\documentclass{article} \usepackage{amsmath} \usepackage{hhline}
\usepackage{graphicx} \usepackage[T2A]{fontenc}
\usepackage[cp1251]{inputenc} \usepackage[russian]{babel}

\begin{document}          \parindent=13pt
\def\ds{\displaystyle}    \def\ss{\scriptsize}
\def\hh{\hskip 1pt}       \def\hs{\hskip 2pt}   \def\h{\hskip 0.2mm}
\def\pr{\prime}
\newcommand{\fbR}{\hbox{\ss\bf R}}
\newcommand{\fbr}{\hbox{\footnotesize\bf r}}
\newcommand{\fbk}{\hbox{\footnotesize\bf k}}
\newcommand{\Tr}{\mathop{\rm T\h r}\nolimits}

\setcounter{page}{1}

\phantom{x}\vskip 25mm

\centerline{\large\bf ANISOTROPY AND SUPERCONDUCTIVITY } \vskip 5mm

\centerline{\large Boris V. Bondarev } \vskip 5mm

\centerline{\it Moscow Aviation Institude, Volokolamskoye Shosse 4,
125871, Moscow, Russia } \vskip 1mm

\centerline{E-mail: bondarev.b@mail.ru} \vskip 2mm

\begin{quote} \hskip 5mm \small The mean field method is applied for
analysis of valence electrons in metals. It is shown that at low
temperatures electrons have two wave-vector distribution patterns.
Isotropic distribution refers to the first pattern. Anisotropic
distribution refers to another pattern, particularly to specific
wave-vector values occurred nearby the Fermi sphere. It is shown
that it is the anisotropy that makes the metal obtain its specific
superconductor features. \end{quote}

\vskip 3mm \centerline{\bf 1. Introduction } \vskip 2mm

\par Prior to discussing the matters disclosed in scope of this
document, I will give my account of the new model Hamiltonian, the
Nobel Prize Winner, P. W. Anderson addressed to in his lecture. This
Hamiltonian is now named, as Anderson Hamiltonian. Hereunder, I'd
like to give some extracts from the above [1, 2].

\par "To describe the case, I put forward the model Hamiltonian,
which is now referred to as Hubbard Hamiltonian: $$
H=\sum\limits_{i\hs j}\hh b_{\hh ij}\hs c_{\hh i\sigma}^{+}\hs
c_{\hh j\sigma}+ \sum\limits_{i}\hh U\hs n_{\hh i\uparrow}\hs n_{\hh
i\downarrow}\hs , $$ where $b_{\hh ij}$ describes the $i$ -- to --
$j$ node electron jump amplitude and $U$ -- repulsion energy of two
single-centered opposite-spin electrons. Need to say, that no
single-centered parallel-spin states are possible.

\par It is assumed that there is a local atomic spin denoted by
${\bf S}$ and existent from the God. It is also assumed that this
kind of a spin interacts with free electrons on their exchangeable
basis. This very statement is exactly referred to Kondo Hamiltonian:
$$ H=\sum\limits_{k\hh\sigma}\hh\varepsilon_k\hs n_{k\hh\sigma}+J\hs
{\bf S}\hh {\bf s}\hs , $$ where $J$ is an empirically supported
exchange integral, $$ {\bf s}=\sum\limits_{k\hh
k^\pr\sigma\sigma^\pr}\hh
c_{k\sigma}^{+}\hh{\bf\sigma}_{\sigma\sigma^\pr}\hh
c_{k^\pr\sigma^\pr}\hs $$ -- local spin density of impurity
conductivity electrons.

\par Anderson model is a very simple one to describe the aforesaid
electronic mechanism. We introduce critically important
single-centered $U$ energy and make specific description of the
impurity atom by applying additional orbital $\varphi_d$ with its
occupation number $n_d$ and birth operator $c_{d\sigma}^{+}$ matched
it. Model Hamiltonian is formulated as follows: $$
H=\sum\limits_{k\hh\sigma}\hh\varepsilon_k\hs n_{k\hh\sigma} + U\hs
n_{\hh d\uparrow}\hs n_{\hh d\downarrow}+E_d\hs (n_{\hh
d\uparrow}+n_{\hh d\downarrow})+\sum\limits_{k\hh\sigma}\hh V_{d\hh
k}\hs c_{d\sigma}^{+}\hh c_{k\sigma}\hs , $$ where, apart from free
electrons and magnetic term with constant $U$, we have added
tunneling $V_{dk}$ term $d - k$ which describes the effect of
electron tunneling through the centrifugal barrier to convert
orbital $\varphi_d$ to one of the Friedel resonances.

\par Before proceeding to other matters of this discussion, I would
like to say some words about simplicity of the model which is
actually more likely superficial than true. Any skills of modeling
result in ability to give up actual and existent, but not
significant aspects of the problem to be of risks for both the
author and reader. As a matter of fact, the author can give up
anything of importance; the reader to be in possession of
excessively delicate experimental techniques or to be excessively
punctual in calculations may oversimplify the conceptualized model
which is chiefly targeted to demonstration of any specific
potential.

\par As referred to the science style, one of my profound beliefs is
that quantum mechanics and statistical physics are in principal so
simple to make many facts, the Nature present with, look like
improbable and just the only demonstration of the reasonable
interpretable mechanism leaves no doubt in adequacy of the
interpretation. Moreover, it concerns the cases when any unexpected
phenomena are correctly predicted; particularly those related to
low-density moment life-span correlation, previously described
orbital-moment quenching, as respects to all the $d$ -- level
impurities, or negative free-electron exchange polarization subject
to further discussion. More often than not, such simplified model
clears up the real nature of the phenomenon, and thereby any 'ab
initio' value calculated for the purpose of specific cases, which,
if they are actually correct, often contain a great number of
elements that more likely conceal than disclose the truth. Thus, the
possibility of providing precise calculation or measurement of
something may be more likely an obstacle than an advantage, since
anything under measurement or calculation occurs to be unessential
from the standpoint of phenomenon mechanism investigation. After
all, any ideal calculation just copies Nature, but not explains it".

\par How it comes that Hamiltonian is named the model one? Because
it is nearly phenomenological - i.e. it is not precisely calculable.
And if there are a few opportunities to make any calculations, then
it is fine. This Hamiltonian may be applied for describing the
experiment under study.

\vskip 5mm \centerline{\bf 2. Anisotropy } \vskip 2mm

\par Now, it is time to describe isotropic distribution [3]. We have
prescribed function $f=f(\bf a)$, i.e. value $f$ depends on vector
$\bf a$. If value $f$ depends on modulus $a$ of this vector only,
the distribution concerned is called isotropic, i.e. it may be
formulated as $f=f(a)$. This kind of isotropy may be represented
graphically (see Fig. 1). We will plot a sphere of radius $a$
centered in the origin of coordinates. So, value $f$ will remain
equal at any point of this sphere, providing that $f=f(\bf a)$ is
the isotropic function. Any other $f=f(\bf a)$ function will be
referred to the anisotropy one.

\unitlength=1mm \centerline{\begin{picture}(65,54)
\put(27,20){\circle*{0.7}}\put(27,20){\vector(-1,1){12.8}}
\put(20.2,27.3){$a$}\put(44,30){$f({\bf a}) = {\rm const}$}
\put(27,20){\vector(-1,-1){19}}\put(3,1.2){$a_{\hh x}$}
\put(27,20){\vector(1,0){27}}\put(52,16){$a_{\hh y}$}
\put(27,20){\vector(0,1){27}}\put(22.3,45){$a_{\hh z}$}
\put(27,20){\unitlength=1.2mm{\special{em:linewidth 0.3pt}}
\put(15,0){\special{em:moveto}} \put(14.94,1.31)
{\special{em:lineto}} \put(14.78,2.61) {\special{em:lineto}}
\put(14.49,3.89) {\special{em:lineto}} \put(14.10,5.13)
{\special{em:lineto}} \put(13.59,6.35) {\special{em:lineto}}
\put(12.99,7.5)  {\special{em:lineto}} \put(12.29,8.61)
{\special{em:lineto}} \put(11.49,9.65) {\special{em:lineto}}
\put(10.61,10.61){\special{em:lineto}} \put(9.65,11.49)
{\special{em:lineto}} \put(8.61,12.29) {\special{em:lineto}}
\put(7.5,12.99)  {\special{em:lineto}} \put(6.35,13.59)
{\special{em:lineto}} \put(5.13,14.10) {\special{em:lineto}}
\put(3.89,14.49) {\special{em:lineto}} \put(2.61,14.78)
{\special{em:lineto}} \put(1.31,14.94) {\special{em:lineto}}
\put(0,15)       {\special{em:lineto}} \put(0,15)
{\special{em:moveto}} \put(-1.31,14.94) {\special{em:lineto}}
\put(-2.61,14.78) {\special{em:lineto}} \put(-3.89,14.49)
{\special{em:lineto}} \put(-5.13,14.10) {\special{em:lineto}}
\put(-6.35,13.59) {\special{em:lineto}} \put(-7.5,12.99)
{\special{em:lineto}} \put(-8.61,12.29) {\special{em:lineto}}
\put(-9.65,11.49) {\special{em:lineto}}
\put(-10.61,10.61){\special{em:lineto}} \put(-11.49,9.65)
{\special{em:lineto}} \put(-12.29,8.61) {\special{em:lineto}}
\put(-12.99,7.5)  {\special{em:lineto}} \put(-13.59,6.35)
{\special{em:lineto}} \put(-14.10,5.13) {\special{em:lineto}}
\put(-14.49,3.89) {\special{em:lineto}} \put(-14.78,2.61)
{\special{em:lineto}} \put(-14.94,1.31) {\special{em:lineto}}
\put(-15,0)       {\special{em:lineto}} \put(-15,0)
{\special{em:moveto}} \put(-14.94,-1.31) {\special{em:lineto}}
\put(-14.78,-2.61) {\special{em:lineto}} \put(-14.49,-3.89)
{\special{em:lineto}} \put(-14.10,-5.13) {\special{em:lineto}}
\put(-13.59,-6.35) {\special{em:lineto}} \put(-12.99,-7.5)
{\special{em:lineto}} \put(-12.29,-8.61) {\special{em:lineto}}
\put(-11.49,-9.65) {\special{em:lineto}}
\put(-10.61,-10.61){\special{em:lineto}} \put(-9.65,-11.49)
{\special{em:lineto}} \put(-8.61,-12.29) {\special{em:lineto}}
\put(-7.5,-12.99)  {\special{em:lineto}} \put(-6.35,-13.59)
{\special{em:lineto}} \put(-5.13,-14.10) {\special{em:lineto}}
\put(-3.89,-14.49) {\special{em:lineto}} \put(-2.61,-14.78)
{\special{em:lineto}} \put(-1.31,-14.94) {\special{em:lineto}}
\put(0,-15)        {\special{em:lineto}} \put(0,-15)
{\special{em:moveto}} \put(1.31,-14.94) {\special{em:lineto}}
\put(2.61,-14.78) {\special{em:lineto}} \put(3.89,-14.49)
{\special{em:lineto}} \put(5.13,-14.10) {\special{em:lineto}}
\put(6.35,-13.59) {\special{em:lineto}} \put(7.5,-12.99)
{\special{em:lineto}} \put(8.61,-12.29) {\special{em:lineto}}
\put(9.65,-11.49) {\special{em:lineto}}
\put(10.61,-10.61){\special{em:lineto}} \put(11.49,-9.65)
{\special{em:lineto}} \put(12.29,-8.61) {\special{em:lineto}}
\put(12.99,-7.5)  {\special{em:lineto}} \put(13.59,-6.35)
{\special{em:lineto}} \put(14.10,-5.13) {\special{em:lineto}}
\put(14.49,-3.89) {\special{em:lineto}} \put(14.78,-2.61)
{\special{em:lineto}} \put(14.94,-1.31) {\special{em:lineto}}
\put(15,0)        {\special{em:lineto}} } \end{picture}}

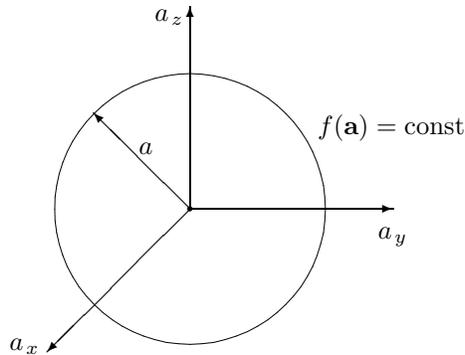
\vskip 3mm \centerline{\it Fig. 1. Isotropic function. } \vskip 3mm

\par Now, we will consider the example of the anisotropic function.
We will plot two vectors. One of them will be an arbitrary vector
$\bf a$ and the other one will be rated as equal, but opposite in
its direction $-\hs\bf a$. Two such vectors are shown in Fig. 2. So,
if it is appeared that function values fail matching in the points
concerned, i.e. $f({\bf a})\ne f({-\hs\bf a})$, this function will
be called the anisotropic one. Some exhaustive examples of the
anisotropic function may be additionally described, but, as a matter
of fact, the information provided is sufficient for understanding.

\unitlength=1mm \centerline{\begin{picture}(65,54)
\put(27,20){\circle*{0.7}}\put(13.8,33.3){\circle*{0.7}}
\put(27,20){\vector(-1,1){13}}\put(10.5,34.5){$\bf a$}
\put(40.3,6.7){\circle*{0.7}}
\put(27,20){\vector(1,-1){13}}\put(37.5,3){$-\hs\bf a$}
\put(35,30){$f({\bf a})\ne f({-\hs\bf a})$}
\put(27,20){\vector(-1,-1){19}}\put(3,1.2){$a_{\hh x}$}
\put(27,20){\vector(1,0){27}}\put(52,16){$a_{\hh y}$}
\put(27,20){\vector(0,1){27}}\put(22.1,45){$a_{\hh z}$}
\end{picture}}

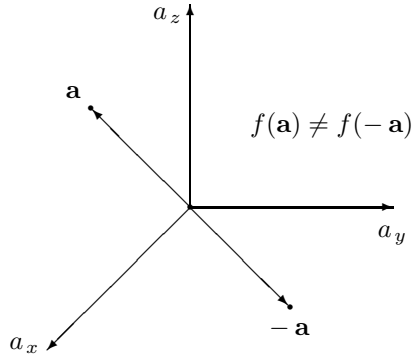
\vskip 3mm \centerline{\it Fig. 2. Example of anisotropic function.
} \vskip 1mm

\vskip 5mm \centerline{\bf 3. Fermi -- Dirac function } \vskip 2mm

\par Now, we will consider formulation of Fermy -- Dirak function
and review its meaning [4]. In the first place, we are chiefly
speaking about electrons taking account to their large and very
large number, for example solid-state body electrons. Secondly,
electrons are subject to Pauli principle, according to which maximum
one electron may be in a particular state. This function has one of
the following simplest formulations: $$ \overline n\hh
(\varepsilon)= \frac{1}{\hs e^{\hh\beta\hh (\varepsilon-\mu
)}+1\hs}\hs , \eqno (3.1) $$ where $\overline n\hh (\varepsilon)$ is
an average number of $\varepsilon$ -- energy electrons,
$\beta=1/{k_B\hs T}$ -- reciprocal temperature, $\mu$ -- chemical
potential. For the function curve, see Fig. 3.

\unitlength=1.5mm \centerline{\begin{picture}(70,38)
\put(0,5){\vector(1,0){70}}\put(68.5,2){$\varepsilon$}
\put(20,5){\vector(0,1){30}}\put(21.4,33){$\overline
n(\varepsilon)$} \put(17.4,6){0}\put(17.4,26){1}
\multiput(35,5)(0,2.07){10}{\line(0,1){1.4}}\put(34,2){$\mu$}
\multiput(35,25)(-1.85,0){19}{\line(-1,0){1.4}}
\put(35,15){\circle*{0.7}}
\multiput(20,15)(1.85,0){8}{\line(1,0){1.4}}
\put(17,14.3){$\frac12$}  \put(35,5)
{\unitlength=1.5mm\special{em:linewidth 0.3pt}
\put(-26,19.88){\special{em:moveto}}
\put(-24,19.84){\special{em:lineto}}
\put(-22,19.76){\special{em:lineto}}
\put(-20,19.65){\special{em:lineto}}
\put(-18,19.47){\special{em:lineto}}
\put(-16,19.21){\special{em:lineto}}
\put(-14,18.85){\special{em:lineto}}
\put(-12,18.33){\special{em:lineto}}
\put(-10,17.62){\special{em:lineto}} \put(
-8,16.64){\special{em:lineto}} \put( -6,15.37){\special{em:lineto}}
\put( -4,13.80){\special{em:lineto}} \put(
-2,11.98){\special{em:lineto}} \put(  0,10.00){\special{em:lineto}}
\put(  2,8.026){\special{em:lineto}} \put(
4,6.200){\special{em:lineto}} \put(  6,4.630){\special{em:lineto}}
\put(  8,3.360){\special{em:lineto}} \put(
10,2.384){\special{em:lineto}} \put( 12,1.664){\special{em:lineto}}
\put( 14,1.147){\special{em:lineto}} \put(
16,0.783){\special{em:lineto}} \put( 18,0.532){\special{em:lineto}}
\put( 20,0.360){\special{em:lineto}} \put(
22,0.243){\special{em:lineto}} \put( 24,0.163){\special{em:lineto}}
\put( 26,0.110){\special{em:lineto}} } \end{picture}}

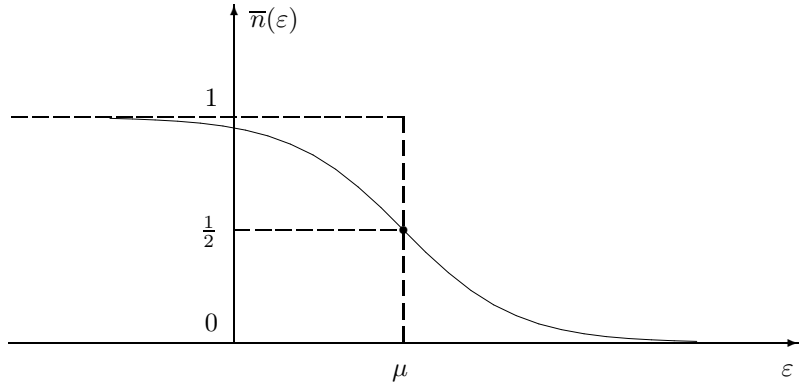
\vskip 1mm \centerline {\it Fig. 3. Fermi -- Dirac function. }
\vskip 2mm

\par Fermi -- Dirac function shall be derived for the purpose of
electron interacting system only. Since electrons interact to each
other, just an approximate equation may be derived for function
$\overline n\hh (\varepsilon)$. The simplest one is mean field
approximation in scope of which it is defined that double
probability is equal to the product of single probabilities. This
kind of approximation is known as statistical independence. As
referred to the mean field approximation, the following equation is
obtained: $$ \overline n\hh (\varepsilon)= \frac{1}{\hs
e^{\hh\beta\hh (\overline\varepsilon-\mu )}+1\hs}\hs , \eqno (3.2)
$$ where $\overline\varepsilon$ is average electron energy which is
equal to the sum of kinetic energy $\varepsilon$ and energy of
interaction of this electron with other electrons.

\par Now, let us remember that in specific representation a
particular state of electron is described by wave function
$\psi_{\hh\fbk n}({\bf r})\hh\chi_\sigma(\xi)$, where $\bf k$ is a
wave vector, $n$ and $\sigma$ -- other numbers, which in combination
with the wave vector make it possible to define the state of
electron. The wave vector is linked with an electron momentum by the
simple relation ${\bf p}=\hbar\hs\bf k$. Now, without going into any
details, we formulate the equation (3.2) as follows: $$
\ln\hs\frac{\hh 1-w_{\hh\fbk}\hh}{w_{\hh\fbk}}= \beta\hs
(\hh\overline\varepsilon_{\fbk}-\mu\hh )\hs , \eqno (3.3) $$ where
$w_{\hh\fbk}=\overline n(\varepsilon_{\fbk})$ is probability of $\bf
k$ vector occupation probability; $\overline\varepsilon_{\fbk}$ is
single electron mean energy; $\varepsilon_{\fbk}$ is electron
kinetic energy. By this means as follows: $$
\overline\varepsilon_{\fbk}=\varepsilon_{\fbk} +
\sum\limits_{\fbk^\pr}\hs\varepsilon_{\fbk\fbk^\pr}\hs
w_{\hh\fbk^\pr}\hs . \eqno (3.4) $$ The addend in this formula is
the mean energy of interaction of an electron with other electrons.
Summand $\varepsilon_{\fbk\fbk^\pr}$ is the energy of interaction of
two electrons with wave vectors $\bf k$ and $\bf k^\pr$.

\vskip 5mm \centerline{\bf 4. Model Hamiltonian } \vskip 2mm

\par Equation (3.3) contains its nonlinearly unknown function
$w_{\hh\fbk}$. Now, for the purpose of the above function, it is
necessary to define the electron interaction energy
$\varepsilon_{\fbk\fbk^\pr}$. Electrons interact with each other
under the Coulomb repulsive potential. But there is rather large
number of ions and other electrons applied for interaction in the
solid body. Need to say that it is very hard to calculate the rate
of electron interaction energy. Therefore, we will use the model
Hamiltonian. We will assume that [5] $$
\varepsilon_{\fbk\fbk^\pr}=I\hs\delta_{\fbk +\fbk^\pr}\hs , \eqno
(4.1) $$ where $\delta_{\fbk +\fbk^\pr}$ is Kronecker delta, $I$ is
$\bf k$ and $\bf k^\pr=-\hs\bf k$ wave vector and electron
interaction energy. As provided in scope of our model, only those
valence electrons are repulsive, which are able to surmount the
crystal area at equal opposite direction velocities.

\par With the model formula applied (4.1) the average electron
energy (3.4) is formulated by the equation as follows: $$
\overline\varepsilon_{\fbk}=\varepsilon_{\fbk} + I\hs
w_{-\hh\fbk}\hs . \eqno (4.2) $$ According to this formula, the
higher $\bf k$ vector electron energy, the more probability $w_{\hh
-\fbk}$ of $-\hh\bf k$ wave vector occupation. Hereby, the $-\hh\bf
k$ wave vector electron somewhat affects the electron forcing it out
of $\bf k$ wave vector state.

\par We will insert formula (4.2) in equation (3.3). We will gain
the following formula: $$ \ln\hs\frac{\hh
1-w_{\hh\fbk}\hh}{w_{\hh\fbk}}= \beta\hs (\hh\varepsilon_{\fbk}+I\hs
w_{-\hh\fbk}-\mu\hh )\hs . \eqno (4.3) $$ Thus, the equation
containing two function values $w_{\hh\fbk}$ and $w_{\hh -\hh\fbk}$
is produced.

\vskip 5mm \centerline{\bf 5. Isotropic and anisotropic electron
distribution } \vskip 2mm

\par If you ask why it has to do with anisotropy, it may be
confirmed by the equation (4.3) which exhibits solution of
anisotropic function $w_{\hh\fbk}$ subject to condition of
$w_{-\hh\fbk}\ne w_{\hh\fbk}$. In this equation we will substitute
vector $\bf k$ for vector $-\hh\bf k$. If to consider that kinetic
energy is the isotropic function, i.e. $\varepsilon_{\hh
-\hh\fbk}=\varepsilon_{\fbk}$, we will formulate the following
equation: $$ \ln\hs\frac{\hh 1-w_{\hh -\fbk}\hh}{w_{\hh -\fbk}}=
\beta\hs (\hh\varepsilon_{\fbk} + I\hs w_{\hh\fbk}-\mu\hh )\hs .
\eqno (5.1) $$ Equations (4.3) and (5.1) produce the system
containing two unknown $w_{\hh\fbk}$ and $w_{\hh -\fbk}$. At the
same time, it is clear enough that probability $w_{\hh\fbk}$ is the
composite $\bf k$ vector function, where electron kinetic energy is
applied as an intervening variable $\varepsilon_{\fbk}$: $$
w_{\hh\fbk}=w(\varepsilon_{\fbk})\hs . \eqno (5.2) $$

\par Combined equations (4.3) and (5.1) exhibit their anisotropic
solution subject to the condition as follows: $$ w_{\hh
-\fbk}=w_{\hh \fbk}\hs . \eqno (5.3) $$ Using this equation we will
eliminate $w_{\hh -\fbk}$ from the combined equations (4.3) and
(5.1). We will find solution of isotropic distribution function by
applying the following equation: $$ \ln\hs\frac{\hh
1-w_{\hh\fbk}\hh}{w_{\hh\fbk}}= \beta\hs (\hh\varepsilon_{\fbk} +
I\hs w_{\hh\fbk}-\mu\hh )\hs . \eqno (5.4) $$

\par There are also some anisotropic distribution functions that
fall out of formula (5.3) when specific wave vector values are
applied: $$ w_{\hh -\fbk}\neq w_{\hh \fbk}\hs . $$ Such kind of
electron state distribution anisotropy may occur even when no
external field is available. While applying forms: $$ w_{\hh
-\fbk}=w_{\hh 1}(\varepsilon_{\hh\fbk})\hs , \hskip 10mm
w_{\hh\fbk}=w_{\hh 2}(\varepsilon_{\hh\fbk})\hs , \eqno (5.5) $$ we
may formulate equations (4.3) and (5.1) by the method as follows: $$
\left.\begin{array}{l} \ds\ln\hs\frac{\hh 1-w_{\hh 1}\hh}{w_{\hh
1}}= \frac{\hh 4\hh}{\tau}\hs(\hh\epsilon + w_{\hh 2}\hh )\hs ,
\medskip \\ \ds\ln\hs\frac{\hh 1-w_{\hh 2}\hh}{w_{\hh 2}}= \frac{\hh
4\hh}{\tau}\hs(\hh\epsilon + w_{\hh 1}\hh )\hs , \\
\end{array}\right\} \eqno (5.6) $$ where $$ \epsilon
=\frac{\hs\varepsilon -\mu}{I}\hs , \hskip 10mm \tau =\frac{\hs
4\hs\theta\hs}{I}\hs . $$ The following functions remain unknown in
the combined equations (5.6): $$ w_{\hh 1}=w_{\hh
1}(\varepsilon)\hskip 10mm\hbox{и}\hskip 10mm w_{\hh 2}=w_{\hh
2}(\varepsilon)\hs . $$

\par If electrons have isotropic wave vector distribution, it is
necessary to insert $w_1=w_2=w_0$ in the combined equation (5.6). In
this case, the equation gained may be formulated by the method as
follows: $$ \epsilon=\frac{\tau}{\hh 4\hh}\hs\ln\hh \frac{\hs
1-w_0\hs}{\hs w_0\hs}-w_0\hs . \eqno (5.7) $$ This equation states
specific dependence of $w_0=w_0(\epsilon)$ with various temperature
values graphically represented in Fig. 4 in the form of
monotonically decreasing curves.

\par If electrons have anisotropic wave vector distribution,
probabilities $w_{\hh 1}$ and $w_{\hh 2}$ in the combined equations
(5.6) shall be considered as various functions $w_{\hh 1}=w_{\hh
1}(\epsilon)$ and $w_{\hh 2}=w_{\hh 2}(\epsilon)$ subject to energy
$\epsilon$. To determine these dependences, we will introduce new
variables $d$ and $s$ applying the relations as follows $$
w_2-w_1=d\hs , \hskip 10mm w_1+w_2=1+s\hs . \eqno (5.8) $$ Without
loss of generality we will assume that nonnegative difference $d$ of
two distribution function values $w_1$ and $w_2$ is $d\geq 0$. At
the same time, $d$ remains equal to $ d\in[\hh 0,\hh 1\hh]$. Value
$s$ may possess the values within the range of $-\hh 1$ to 1: $s\in
[\hh -1,\hh 1\hh]$. We will determine the equalities (5.8), as
regards the probabilities $w_{\hh 1}$ and $w_{\hh 2}$: $$
w_1=\frac{1}{\hh 2\hh}\hs (\hh 1+s-d\hh)\hs , \hskip 10mm
w_2=\frac{1}{\hh 2\hh}\hs (\hh 1+s+d\hh)\hs . \eqno (5.9) $$

\par We will transform the combined equations (5.6) by applying the
formulas (5.9). We will firstly subtract specific equation from one
of the combined equations and then we will add the equations. As a
result, we will obtain the following combined equations: $$
\ds\frac{\hs (\hh 1+d\hh )^2-s^{\hh 2}\hh}{\hs (\hh 1-d\hh
)^2-s^{\hh 2}\hh}= e^{\hh 4\hh d/\tau}\hs , \hskip 10mm \ds
\epsilon=\frac{\tau}{\hh 8\hh}\hs\ln\hh\frac{\hs (\hh 1-s\hh
)^2-d^{\hh 2}\hh} {\hs (\hh 1+s\hh )^2-d^{\hh 2}\hh}-\frac{1}{\hh
2\hh}\hs (\hh 1+s\hh)\hs . \eqno (5.10) $$ The first equation of the
combined ones may be easily resolved against $s$: $$
s(d)=\pm\hs\sqrt{\frac{\hh (\hh 1-d\hh)^{\hh 2}\hs e^{\hh 4\hh
d/\tau}- (\hh 1+d\hh)^{\hh 2}\hh}{e^{\hh 4\hh d/\tau}-1}}\hs . $$

\phantom{x}

\vskip -10mm \centerline{\unitlength=1.2mm\begin{picture}(85,67)
\put(30,41){\it 1}\put(47,17){\it 1} \put(13,44){\it
2}\put(65,12.4){\it 2} \put(4,42){\it 3} \put(75,14){\it 3}
\put(0,10){\vector(1,0){87}}\put(80,6){$\varepsilon-\mu$}
\multiput(20,10.5)(0,2){20}{\line(0,1){1.4}}
\multiput(20,50)(2,0){20}{\line(1,0){1.4}}
\put(20,10){\line(0,-1){1}}\put(17,5){$-I$}
\put(40,30){\circle*{0.7}}
\multiput(40,10.5)(0,2){10}{\line(0,1){1.4}}
\put(40,10){\line(0,-1){1}}\put(37,4.5){$-\frac12\hh I$}
\put(60,10){\line(0,-1){1}}\put(59.2,5){0}
\put(60,10){\vector(0,1){50}}\put(62,58.5){$w$} 
\put(0,50){\line(1,0){21}}\put(62,49){1,0}
\multiput(59.5,30)(-2,0){10}{\line(-1,0){1.4}}
\put(60,30){\line(1,0){1}}\put(62,29){0,5}
\put(60,10){\unitlength=1.2mm\special{em:linewidth
0.3pt}\put(-40,40) {\special{em:moveto}}
\put(0,0){\special{em:lineto}} }
\put(60,10){\unitlength=1.2mm\special{em:linewidth 0.3pt}
\put(11.62,0.3333)  {\special{em:moveto}} \put(9.528,0.6667)
{\special{em:lineto}} \put(8.160,1)  {\special{em:lineto}}
\put(5.360,2)  {\special{em:lineto}} \put(3.280,3)
{\special{em:lineto}} \put(1.492,4)  {\special{em:lineto}}
\put(-0.136,5)  {\special{em:lineto}} \put(-1.664,6)
{\special{em:lineto}} \put(-3.122,7)  {\special{em:lineto}}
\put(-4.536,8)  {\special{em:lineto}} \put(-5.908,9)
{\special{em:lineto}} \put(-7.252,10)  {\special{em:lineto}}
\put(-8.576,11)  {\special{em:lineto}} \put(-9.884,12)
{\special{em:lineto}} \put(-11.17,13)  {\special{em:lineto}}
\put(-12.45,14)  {\special{em:lineto}} \put(-13.72,15)
{\special{em:lineto}} \put(-14.99,16)  {\special{em:lineto}}
\put(-16.24,17)  {\special{em:lineto}} \put(-17.50,18)
{\special{em:lineto}} \put(-18.75,19)  {\special{em:lineto}}
\put(-20.00,20)  {\special{em:lineto}} \put(-21.25,21)
{\special{em:lineto}} \put(-22.50,22)  {\special{em:lineto}}
\put(-23.76,23)  {\special{em:lineto}} \put(-25.01,24)
{\special{em:lineto}} \put(-26.28,25)  {\special{em:lineto}}
\put(-27.55,26)  {\special{em:lineto}} \put(-28.83,27)
{\special{em:lineto}} \put(-30.12,28)  {\special{em:lineto}}
\put(-31.42,29)  {\special{em:lineto}} \put(-32.75,30)
{\special{em:lineto}} \put(-34.09,31)  {\special{em:lineto}}
\put(-35.46,32)  {\special{em:lineto}} \put(-36.88,33)
{\special{em:lineto}} \put(-38.34,34)  {\special{em:lineto}}
\put(-39.86,35)  {\special{em:lineto}} \put(-41.48,36)
{\special{em:lineto}} \put(-43.28,37)  {\special{em:lineto}}
\put(-45.36,38)  {\special{em:lineto}} \put(-48.16,39)
{\special{em:lineto}} \put(-49.52,39.33)  {\special{em:lineto}}
\put(-51.60,39.67)  {\special{em:lineto}} }
\put(60,10){\unitlength=1.2mm\special{em:linewidth 0.3pt}
\put(27.00,1.2)  {\special{em:moveto}}
\put(25.99,1.3){\special{em:lineto}} \put(21.55,2)
{\special{em:lineto}} \put(17.10,3)  {\special{em:lineto}}
\put(13.58,4)  {\special{em:lineto}} \put(10.57,5)
{\special{em:lineto}} \put(7.880,6)  {\special{em:lineto}}
\put(5.408,7)  {\special{em:lineto}} \put(3.088,8)
{\special{em:lineto}} \put(0.8960,9)  {\special{em:lineto}}
\put(-1.208,10)  {\special{em:lineto}} \put(-3.244,11)
{\special{em:lineto}} \put(-5.224,12)  {\special{em:lineto}}
\put(-7.152,13)  {\special{em:lineto}} \put(-9.048,14)
{\special{em:lineto}} \put(-10.91,15)  {\special{em:lineto}}
\put(-12.76,16)  {\special{em:lineto}} \put(-14.58,17)
{\special{em:lineto}} \put(-16.40,18)  {\special{em:lineto}}
\put(-18.20,19)  {\special{em:lineto}} \put(-20.00,20)
{\special{em:lineto}} \put(-21.80,21)  {\special{em:lineto}}
\put(-23.60,22)  {\special{em:lineto}} \put(-25.42,23)
{\special{em:lineto}} \put(-27.24,24)  {\special{em:lineto}}
\put(-29.09,25)  {\special{em:lineto}} \put(-30.95,26)
{\special{em:lineto}} \put(-32.85,27)  {\special{em:lineto}}
\put(-34.78,28)  {\special{em:lineto}} \put(-36.76,29)
{\special{em:lineto}} \put(-38.79,30)  {\special{em:lineto}}
\put(-40.88,31)  {\special{em:lineto}} \put(-43.08,32)
{\special{em:lineto}} \put(-45.40,33)  {\special{em:lineto}}
\put(-47.88,34)  {\special{em:lineto}} \put(-50.56,35)
{\special{em:lineto}} \put(-53.56,36)  {\special{em:lineto}}
\put(-57.08,37)  {\special{em:lineto}} \put(-60.00,37.7)
{\special{em:lineto}} } \end{picture}}

\vskip -2mm \centerline {\it Fig. 4. Isotropic function of
distribution of conductivity electron energy } \centerline {\it at
various temperature values $\tau$: 1 -- $\tau=0$; 2 -- $\tau=0,25$;
3 -- $\tau=0,8$. } \vskip 3mm

\noindent As concerns the relations (5.9), probabilities $w_{\hh 1}$
and $w_{\hh 2}$ may be also considered as $d$ functions: $w_{\hh
1}=w_{\hh 1}(d)$, $w_{\hh 2}=w_{\hh 2}(d)$. With the second equation
of the combined ones (5.10) applied, we may express electron energy
$\epsilon$ in terms of parameter $d$. Using the dependences produced
specific graphs of functions $w_{\hh 1}=w_{\hh 1}(\epsilon)$ and
$w_{\hh 2}=w_{\hh 2}(\epsilon)$ may be easily plotted for various
temperature values. For the plotted curve, see Fig. 5.

\vskip 7mm \centerline{\unitlength=1.2mm\begin{picture}(87,57)
\put(21,11){\it 1}\put(25,16){\it 2}\put(30.5,23.2){\it
3}\put(34,28){\it 4} \put(54,51){\it 1}\put(48.5,44.5){\it 2}
\put(0,10){\vector(1,0){87}}\put(79,5){$\varepsilon-\mu$}
\multiput(20,10.5)(0,2){20}{\line(0,1){1.4}}
\put(20,10){\line(0,-1){1}}\put(17,5){$-I$}
\put(40,30){\circle*{0.7}}\multiput(40,10.5)(0,2){10}{\line(0,1){1.4}}
\put(40,10){\line(0,-1){1}}\put(37,4.5){$-\frac12\hh I$}
\put(60,10){\line(0,-1){1}}\put(59.2,5){0}
\put(60,10){\vector(0,1){50}}\put(62,58.5){$w$} 
\put(20,50){\unitlength=1.2mm\line(1,0){41}}\put(62,49){1,0}
\put(20,50.2){\unitlength=1.2mm\line(1,0){40}}
\put(20,9.8){\unitlength=1.2mm\line(1,0){40}}
\multiput(20,50)(-2,0){10}{\line(-1,0){1.4}}
\multiput(59.5,30)(-2,0){10}{\line(-1,0){1.4}}
\put(60,30){\line(1,0){1}}\put(62,29){0,5}
\put(60,10){\unitlength=1.2mm\special{em:linewidth 0.3pt}
\put(-43.84,36.59)  {\special{em:moveto}} \put(-43.76,35.77)
{\special{em:lineto}} \put(-43.60,34.84)  {\special{em:lineto}}
\put(-43.40,33.82)  {\special{em:lineto}} \put(-43.16,32.73)
{\special{em:lineto}} \put(-42.88,31.57)  {\special{em:lineto}}
\put(-40.96,23.96)  {\special{em:lineto}} \put(-40.64,22.64)
{\special{em:lineto}} \put(-40.32,21.32)  {\special{em:lineto}}
\put(-39.98,19.99)  {\special{em:lineto}} \put(-39.66,18.66)
{\special{em:lineto}} \put(-39.32,17.33)  {\special{em:lineto}}
\put(-38.98,16.00)  {\special{em:lineto}} \put(-38.64,14.66)
{\special{em:lineto}} \put(-38.26,13.33)  {\special{em:lineto}}
\put(-37.88,12.00)  {\special{em:lineto}} \put(-37.47,10.66)
{\special{em:lineto}} \put(-37.02,9.336)  {\special{em:lineto}}
\put(-36.54,8.000)  {\special{em:lineto}} \put(-35.98,6.668)
{\special{em:lineto}} \put(-35.32,5.336)  {\special{em:lineto}}
\put(-34.51,4.000)  {\special{em:lineto}} \put(-33.41,2.664)
{\special{em:lineto}} \put(-32.06,1.600)  {\special{em:lineto}}
\put(-31.57,1.336)  {\special{em:lineto}} \put(-30.27,0.800)
{\special{em:lineto}} \put(-29.20,0.536)  {\special{em:lineto}}
\put(-27.56,0.264)  {\special{em:lineto}} \put(-1.100,0.004)
{\special{em:lineto}} \put(-0.344,0.004)  {\special{em:lineto}}
\put(0.3148,0.020)  {\special{em:lineto}} \put(0.9626,0.044)
{\special{em:lineto}} \put(1.6024,0.096)  {\special{em:lineto}}
\put(2.2572,0.208)  {\special{em:lineto}} \put(2.8704,0.428)
{\special{em:lineto}} \put(3.4012,0.845)  {\special{em:lineto}}
\put(3.7720,1.567)  {\special{em:lineto}} \put(3.8710,3.401)
{\special{em:lineto}} \put(3.7714,4.232)  {\special{em:lineto}}
\put(3.6118,5.162)  {\special{em:lineto}} \put(3.4014,6.181)
{\special{em:lineto}} \put(3.1508,7.275)  {\special{em:lineto}}
\put(2.8701,8.431)  {\special{em:lineto}} \put(2.5689,9.634)
{\special{em:lineto}} \put(2.2545,10.88)  {\special{em:lineto}}
\put(1.9330,12.14)  {\special{em:lineto}} \put(1.6077,13.43)
{\special{em:lineto}} \put(1.2823,14.73)  {\special{em:lineto}}
\put(0.9579,16.04)  {\special{em:lineto}} \put(0.6335,17.36)
{\special{em:lineto}} \put(0.3093,18.69)  {\special{em:lineto}}
\put(-0.014,20.01)  {\special{em:lineto}} \put(-0.344,21.34)
{\special{em:lineto}} \put(-0.669,22.67)  {\special{em:lineto}}
\put(-1.016,24.00)  {\special{em:lineto}} \put(-1.401,25.34)
{\special{em:lineto}} \put(-1.729,26.67)  {\special{em:lineto}}
\put(-2.127,28.00)  {\special{em:lineto}} \put(-2.523,29.33)
{\special{em:lineto}} \put(-2.899,30.67)  {\special{em:lineto}}
\put(-3.456,32.00)  {\special{em:lineto}} \put(-3.993,33.33)
{\special{em:lineto}} \put(-4.607,34.67)  {\special{em:lineto}}
\put(-5.484,36.00)  {\special{em:lineto}} \put(-6.555,37.33)
{\special{em:lineto}} \put(-7.088,37.87)  {\special{em:lineto}}
\put(-8.390,38.67)  {\special{em:lineto}} \put(-9.074,38.93)
{\special{em:lineto}} \put(-10.68,39.47)  {\special{em:lineto}}
\put(-12.56,39.73)  {\special{em:lineto}} \put(-38.98,40.00)
{\special{em:lineto}} \put(-39.66,40.00)  {\special{em:lineto}}
\put(-40.32,39.98)  {\special{em:lineto}} \put(-40.96,39.96)
{\special{em:lineto}} \put(-41.60,39.90)  {\special{em:lineto}}
\put(-42.24,39.79)  {\special{em:lineto}} \put(-42.88,39.57)
{\special{em:lineto}} \put(-43.40,39.16)  {\special{em:lineto}}
\put(-43.76,38.43)  {\special{em:lineto}} \put(-43.84,36.59)
{\special{em:lineto}} }
\put(60,10){\unitlength=1.2mm\special{em:linewidth 0.3pt}
\put(-36.61,28.25)  {\special{em:moveto}} \put(-36.56,27.54)
{\special{em:lineto}} \put(-36.49,26.81)  {\special{em:lineto}}
\put(-36.30,26.00)  {\special{em:lineto}} \put(-36.16,25.22)
{\special{em:lineto}} \put(-35.90,24.36)  {\special{em:lineto}}
\put(-35.60,23.46)  {\special{em:lineto}} \put(-35.30,22.58)
{\special{em:lineto}} \put(-34.92,21.62)  {\special{em:lineto}}
\put(-34.49,20.64)  {\special{em:lineto}} \put(-34.04,19.66)
{\special{em:lineto}} \put(-33.52,18.62)  {\special{em:lineto}}
\put(-32.93,17.55)  {\special{em:lineto}} \put(-32.27,16.47)
{\special{em:lineto}} \put(-31.54,15.34)  {\special{em:lineto}}
\put(-30.70,14.16)  {\special{em:lineto}} \put(-29.75,12.97)
{\special{em:lineto}} \put(-28.64,11.70)  {\special{em:lineto}}
\put(-27.28,10.36)  {\special{em:lineto}} \put(-25.57,8.904)
{\special{em:lineto}} \put(-23.82,7.696)  {\special{em:lineto}}
\put(-22.74,7.064)  {\special{em:lineto}} \put(-20.56,6.016)
{\special{em:lineto}} \put(-17.259,4.936)  {\special{em:lineto}}
\put(-13.507,4.348)  {\special{em:lineto}} \put(-11.362,4.300)
{\special{em:lineto}} \put(-9.7588,4.428)  {\special{em:lineto}}
\put(-8.4684,4.664)  {\special{em:lineto}} \put(-7.3924,4.988)
{\special{em:lineto}} \put(-6.4840,5.376)  {\special{em:lineto}}
\put(-5.7180,5.848)  {\special{em:lineto}} \put(-5.0736,6.376)
{\special{em:lineto}} \put(-4.5384,6.984)  {\special{em:lineto}}
\put(-4.1092,7.640)  {\special{em:lineto}} \put(-3.7788,8.388)
{\special{em:lineto}} \put(-3.5472,9.188)  {\special{em:lineto}}
\put(-3.4080,10.12)  {\special{em:lineto}} \put(-3.379, 11.74)
{\special{em:lineto}} \put(-3.4399,12.46)  {\special{em:lineto}}
\put(-3.5434,13.20)  {\special{em:lineto}} \put(-3.6889,13.99)
{\special{em:lineto}} \put(-3.8770,14.80)  {\special{em:lineto}}
\put(-4.1085,15.65)  {\special{em:lineto}} \put(-4.3842,16.53)
{\special{em:lineto}} \put(-4.7046,17.43)  {\special{em:lineto}}
\put(-5.0727,18.38)  {\special{em:lineto}} \put(-5.4901,19.35)
{\special{em:lineto}} \put(-5.9590,20.35)  {\special{em:lineto}}
\put(-6.4850,21.38)  {\special{em:lineto}} \put(-7.0731,22.44)
{\special{em:lineto}} \put(-7.7300,23.54)  {\special{em:lineto}}
\put(-8.4674,24.67)  {\special{em:lineto}} \put(-9.3005,25.83)
{\special{em:lineto}} \put(-10.252,27.04)  {\special{em:lineto}}
\put(-11.364,28.30)  {\special{em:lineto}} \put(-12.705,29.63)
{\special{em:lineto}} \put(-14.444,31.10)  {\special{em:lineto}}
\put(-15.347,31.76)  {\special{em:lineto}} \put(-16.176,32.30)
{\special{em:lineto}} \put(-17.259,32.94)  {\special{em:lineto}}
\put(-19.453,33.99)  {\special{em:lineto}} \put(-22.74,35.06)
{\special{em:lineto}} \put(-26.49,35.65)  {\special{em:lineto}}
\put(-28.64,35.70)  {\special{em:lineto}} \put(-30.24,35.57)
{\special{em:lineto}} \put(-31.54,35.34)  {\special{em:lineto}}
\put(-32.61,35.01)  {\special{em:lineto}} \put(-33.52,34.62)
{\special{em:lineto}} \put(-34.28,34.15)  {\special{em:lineto}}
\put(-34.92,33.62)  {\special{em:lineto}} \put(-35.45,33.02)
{\special{em:lineto}} \put(-35.90,32.36)  {\special{em:lineto}}
\put(-36.23,31.61)  {\special{em:lineto}} \put(-36.49,30.81)
{\special{em:lineto}} \put(-36.55,29.88)  {\special{em:lineto}}
\put(-36.61,28.25)  {\special{em:lineto}} }
\put(60,10){\unitlength=1.2mm\special{em:linewidth 0.3pt}
\put(-21.357,28.06)  {\special{em:moveto}} \put(-23.367,28.62)
{\special{em:lineto}} \put(-24.498,28.62)  {\special{em:lineto}}
\put(-25.346,28.60)  {\special{em:lineto}} \put(-26.022,28.47)
{\special{em:lineto}} \put(-26.574,28.24)  {\special{em:lineto}}
\put(-27.053,27.95)  {\special{em:lineto}} \put(-27.446,27.69)
{\special{em:lineto}} \put(-27.779,27.38)  {\special{em:lineto}}
\put(-28.059,27.04)  {\special{em:lineto}} \put(-28.299,26.66)
{\special{em:lineto}} \put(-28.460,26.28)  {\special{em:lineto}}
\put(-28.609,25.78)  {\special{em:lineto}} \put(-28.722,25.34)
{\special{em:lineto}} \put(-28.756,24.74)  {\special{em:lineto}}
\put(-28.756,23.74)  {\special{em:lineto}} \put(-28.722,23.34)
{\special{em:lineto}} \put(-28.609,22.78)  {\special{em:lineto}}
\put(-28.460,22.28)  {\special{em:lineto}} \put(-28.299,21.66)
{\special{em:lineto}} \put(-28.059,21.04)  {\special{em:lineto}}
\put(-27.779,20.38)  {\special{em:lineto}} \put(-27.446,19.69)
{\special{em:lineto}} \put(-27.053,18.95)  {\special{em:lineto}}
\put(-26.574,18.24)  {\special{em:lineto}} \put(-26.022,17.47)
{\special{em:lineto}} \put(-25.346,16.60)  {\special{em:lineto}}
\put(-24.498,15.62)  {\special{em:lineto}} \put(-23.367,14.62)
{\special{em:lineto}} \put(-21.357,13.06)  {\special{em:lineto}}
\put(-18.642,11.94)  {\special{em:lineto}} \put(-16.633,11.38)
{\special{em:lineto}} \put(-15.502,11.38)  {\special{em:lineto}}
\put(-14.653,11.40)  {\special{em:lineto}} \put(-13.979,11.54)
{\special{em:lineto}} \put(-13.427,11.76)  {\special{em:lineto}}
\put(-12.947,12.05)  {\special{em:lineto}} \put(-12.555,12.31)
{\special{em:lineto}} \put(-12.222,12.62)  {\special{em:lineto}}
\put(-11.941,12.96)  {\special{em:lineto}} \put(-11.700,13.34)
{\special{em:lineto}} \put(-11.541,13.72)  {\special{em:lineto}}
\put(-11.392,14.22)  {\special{em:lineto}} \put(-11.277,14.66)
{\special{em:lineto}} \put(-11.244,15.26)  {\special{em:lineto}}
\put(-11.244,16.26)  {\special{em:lineto}} \put(-11.277,16.66)
{\special{em:lineto}} \put(-11.392,17.22)  {\special{em:lineto}}
\put(-11.541,17.72)  {\special{em:lineto}} \put(-11.700,18.34)
{\special{em:lineto}} \put(-11.941,18.94)  {\special{em:lineto}}
\put(-12.222,19.62)  {\special{em:lineto}} \put(-12.555,20.31)
{\special{em:lineto}} \put(-12.947,21.05)  {\special{em:lineto}}
\put(-13.427,21.76)  {\special{em:lineto}} \put(-13.979,22.54)
{\special{em:lineto}} \put(-14.653,23.40)  {\special{em:lineto}}
\put(-15.502,24.38)  {\special{em:lineto}} \put(-16.633,25.38)
{\special{em:lineto}} \put(-18.642,26.94)  {\special{em:lineto}}
\put(-21.357,28.06)  {\special{em:lineto}} } \end{picture}}

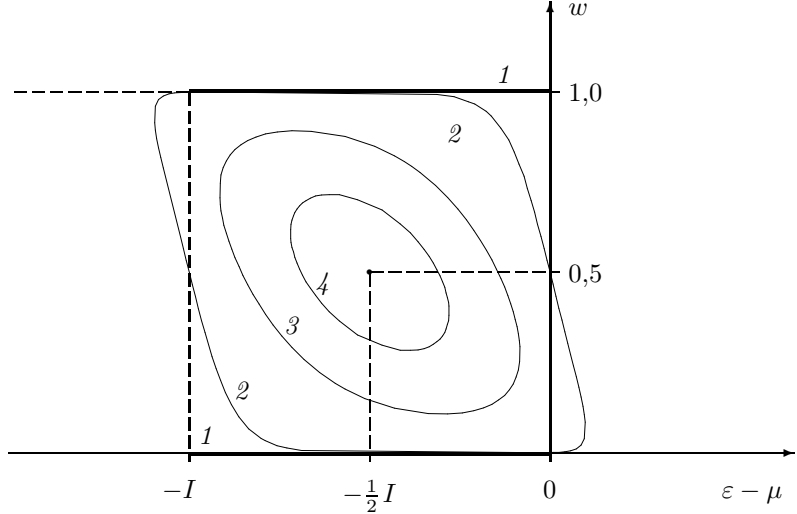
\vskip -2mm \centerline {\it Fig. 5. Anisotropic function of
distribution of conductivity electron energy } \centerline {\it at
various temperature values $\tau$: 1 -- $\tau=0$; 2 -- $\tau=0,25$;
3 -- $\tau=0,8$; 4 -- $\tau=0,95$. }\vskip 2mm

\par The pattern of distribution of electrons by their states
depends on their relation between metal temperature $T$ and critical
temperature: $$ T_c=\frac{I}{\hs 4\hs k_B\hh}\hs . $$ At the
temperature of $T\geq T_c$ the distribution function $w=w(\epsilon)$
is single-valued and satisfies the condition (5.3), as respects all
the $\epsilon$ energy values. At the temperature of $T<T_c$ the
energy is limited by $(\epsilon_1,\hh\epsilon_2)$ with function
$w=w(\epsilon)$ possessing any of three values at every point of the
limit, particularly $w_1(\epsilon)<w_0(\epsilon)<w_2(\epsilon)$.
Being out of the aforesaid limit, the distribution function
$w=w(\epsilon)$ possesses only a single value $w_0(\epsilon)$. Thus,
equation (5.4) is resolved by applying function
$w_{\hh\fbk}=w_0(\epsilon_{\hh\fbk})$ to describe isotropic wave
vector electron distribution.

\vskip 5mm \centerline{\bf 6. Order parameter } \vskip 2mm

\par At $T<T_c$ some kind of anisotropic wave vector electron
distribution may occur in the narrow layer $S$ under Fermi surface
$\varepsilon_{\fbk}=\mu$. This kind of distribution is formulated by
$$ w_{\hh\fbk}=w_2(\epsilon_{\hh\fbk})\hs , \hskip 10mm
 w_{\hh -\fbk}=w_1(\epsilon_{\hh\fbk}) \eqno (6.1) $$ subject to
$\epsilon\in (\epsilon_1,\hh\epsilon_2)$. Difference $d=w_{\hh
2}-w_{\hh 1}$ of two anisotropic electron distribution function
values possesses the largest value $d_{\hh max}$ subject to
$\epsilon=0,5$. In this case, $w_0=0,5$ and $s=0$. We will determine
difference $d_{\hh max}$ from temperature $\tau$ by applying $s=0$
in the first equation (5.10): $$ \frac{\hs 2\hs d_{\hh
max}\hs}{\tau} =\ln\hh\frac{\hs 1+d_{\hh max}\hh} {\hs 1-d_{\hh
max}\hh}\hs . \eqno (6.2) $$ For dependence curve, see Fig. 6.

\vskip 10mm \unitlength=1mm \centerline{\begin{picture}(79,57)
\put(12,10){\vector(1,0){59.5}}\put(69.5,5){$\tau$}
\put(12,10){\line(0,-1){1}}\put(11.2,5){0}
\put(32,10){\line(0,-1){1}}\put(29.7,5){0,5}
\put(52,10){\line(0,-1){1}}\put(51.2,5){1}
\put(12,10){\vector(0,1){50}}\put(14,57.4){$d_{\hh max} (\tau)$}
\put(12,10){\line(-1,0){1}}\put(4,9){$0$}
\put(12,30){\line(-1,0){1}}\put(4,29){$0,5$}
\put(12,50){\line(-1,0){1}}\put(4,49){$1,0$}
\put(12,10){\unitlength=1mm\special{em:linewidth 0.3pt}
\put(40.000,0)  {\special{em:moveto}} \put(39.967,2)
{\special{em:lineto}} \put(39.886,4)  {\special{em:lineto}}
\put(39.698,6)  {\special{em:lineto}} \put(39.461,8)
{\special{em:lineto}} \put(39.152,10)  {\special{em:lineto}}
\put(38.770,12)  {\special{em:lineto}} \put(38.310,14)
{\special{em:lineto}} \put(37.767,16)  {\special{em:lineto}}
\put(37.136,18)  {\special{em:lineto}} \put(36.410,20)
{\special{em:lineto}} \put(35.577,22)  {\special{em:lineto}}
\put(34.625,24)  {\special{em:lineto}} \put(33.536,26)
{\special{em:lineto}} \put(32.284,28)  {\special{em:lineto}}
\put(30.834,30)  {\special{em:lineto}} \put(29.128,32)
{\special{em:lineto}} \put(27.067,34)  {\special{em:lineto}}
\put(24.453,36)  {\special{em:lineto}} \put(23.830,36.40)
{\special{em:lineto}} \put(23.159,36.80)  {\special{em:lineto}}
\put(22.431,37.20)  {\special{em:lineto}} \put(21.634,37.60)
{\special{em:lineto}} \put(20.745,38)  {\special{em:lineto}}
\put(19.734,38.40)  {\special{em:lineto}} \put(18.544,38.80)
{\special{em:lineto}} \put(17.062,39.20)  {\special{em:lineto}}
\put(16.76,39.20)  {\special{em:lineto}} \put(14.80,39.60)
{\special{em:lineto}} \put(13.20,39.80)  {\special{em:lineto}}
\put(10.50,39.96)  {\special{em:lineto}} \put(0,40)
{\special{em:lineto}} } \end{picture}}

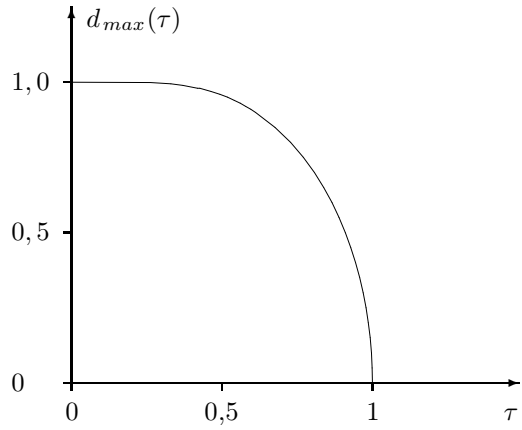
\vskip -2mm \centerline {\it Fig. 6. Electron distribution
anisotropy parameter $d_{\hh max}$, as $\tau$ temperature function.
} \vskip 2mm

\vskip 5mm \centerline{\bf 7. Electron $T=0$ distribution } \vskip
2mm

\par At $T=0$ the isotropic distribution function is formulated as
follows: $$ \hskip 34mm w_{\hh\fbk}=\left\{\begin{array}{ccl} 1 &
\hbox{at} & \varepsilon_{\hh\fbk}\leq\mu - I\hs , \medskip \\
-\hs\ds\frac{1}{\hh I\hh}\hs (\hh\varepsilon_{\hh\fbk}-\mu\hh ) &
\hbox{at} & \mu -I<\varepsilon_{\hh\fbk}<\mu\hs , \hskip 34mm (7.1)
\medskip \\ 0 & \hbox{at} & \varepsilon_{\hh\fbk}\geq\mu\hs .
\end{array}\right. $$ As for the anisotropic distribution, it is
formulated as follows: $$ w_{\hh\fbk}=1 \hskip 4mm \hbox{at} \hskip
4mm \varepsilon_{\hh\fbk}\leq\mu - I\hs , $$ $$ w_{\hh\fbk}=1\hs ,
\hskip 2mm w_{\hh -\fbk}=0 \hskip 7mm \hbox{or} \hskip 7mm
w_{\hh\fbk}=0\hs , \hskip 2mm w_{\hh -\fbk}=1\hskip 7mm \eqno (7.2)
$$ $$ \hbox{at} \hskip 4mm \mu -I<\varepsilon_{\hh\fbk}<\mu\hs ,
\eqno (7.3) $$ $$ w_{\hh\fbk}=0 \hskip 4mm \hbox{at} \hskip 4mm
\varepsilon_{\hh\fbk}\geq\mu\hs . $$ As provided by the formula
(7.2), layer $S$ may be determined under Fermi surface by the
inequality (7.3), in which the electrons have anisotropic wave
vector distribution, i.e. one of the both $\bf k$ and $-\hh\bf k$
wave vector states in this layer is free and another one is
occupied. For the function curves, see Fig. 7. Apparently, electron
distribution function obtains its three $S$ layer values. And what
is the matter it stands for? The answer is in the value of energy,
the isotropic or anisotropic distribution electrons exhibit. The
electrons gain their steady state when they have the lowest energy.

\vskip -2mm \centerline{\unitlength=1.2mm\begin{picture}(87,67)
\put(0,10){\vector(1,0){87}}\put(79,5){$\varepsilon-\mu$}
\multiput(20,10.5)(0,2){20} {\line(0,1){1.4}}
\put(20,10){\line(0,-1){1}}\put(17,5){$-I$}
\put(40,10){\line(0,-1){1}}
\put(60,10){\line(0,-1){1}}\put(59.2,5){0}
\put(60,10){\vector(0,1){50}}\put(62,58.5){$w$}
\put(0,50){\unitlength=1.2mm\line(1,0){61}}\put(62,49){1,0}
\put(39,12){\it 2}\put(39,52){\it 2}
\put(0,50.2){\unitlength=1.2mm\line(1,0){60}}
\put(20,9.8){\unitlength=1.2mm\line(1,0){60}} \put(39,32){\it 1}
\put(60,9.8){\unitlength=1.2mm\line(-1,1){40}}
\put(60,10){\unitlength=1.2mm\line(-1,1){40}}
\put(60,30){\line(1,0){1}}\put(62,29){0,5} \end{picture}}\vskip -2mm

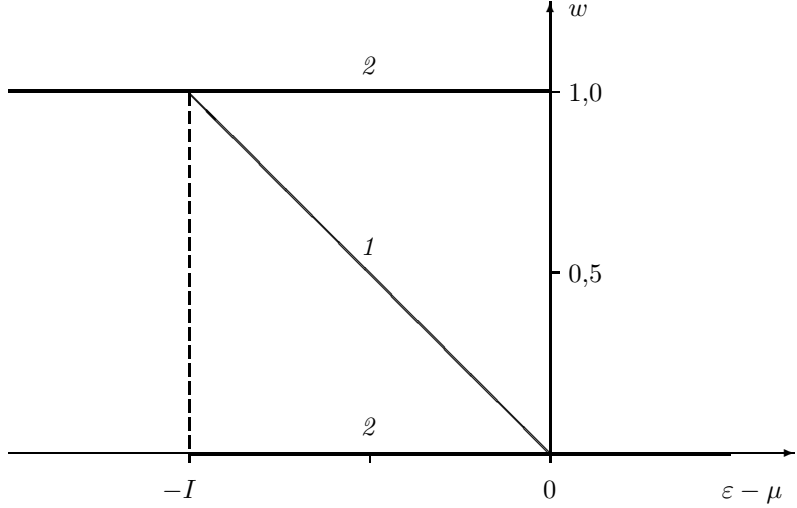
\hskip -13pt {\it Fig. 7. Isotropic and anisotropic distribution of
conductivity electrons depending on their kinetic energy }
\centerline {\it at temperature $\tau=0$: 1 -- isotropic
distribution, 2 -- anisotropic distribution }

\vskip 5mm \centerline{\bf 8. Superconductivity. Energy of states }
\vskip 2mm

\par In scope of normalization conditions, the average itinerant
electron velocity may be defined by the formula as follows: $$ {\bf
v}=\frac{\hs G\hs\hbar\hs}{\hs m\hs\overline N\hs}\hs
\sum\limits_{\fbk}\hs{\bf k}\hs w_{\hh\fbk}\hs , \eqno (8.1) $$
where $G=2\hh n_s$ is a number of states in one node and $\overline
N$ -- mean number of conductivity electrons in a crystal. If the
distribution function is isotropic, mean electron velocity ${\bf v}$
gets equal to zero. Formula (8.1) may assign specific nonzero
electron ordered motion velocity values to some anisotropic
distribution functions, i.e. these distribution functions are
applicable for defining electric current. If there are steady-state
currents to exist with no external fields available, than such
itinerant electron system states shall be considered as the
superconductive ones [6-8].

\par We will assume that the state of electron gas is described by
the anisotropic distribution function (6.1) or (7.2). In this case,
mean electron ordered motion velocity modulus $\bf v$ may assign any
value rated from zero to certain $v_{\hh max}$. The mean velocity
will be equal to zero, providing that free pairs and those occupied
by wave vectors $\bf k$ and $\bf k$ are chaotically distributed
within layer $S$. If all the states concerned are occupied in one
half of the layer (this is to say at $k_x>0$) and free in another
half of the layer (at $k_x<0$), the electrons will gain their
maximum ordered motion velocity. The value assigned by the mean
electron velocity is defined by the nature of initial electron gas
state. If the pattern of anisotropic wave vector electron
distribution is rather steady with respect to small environment
variations, the electron velocity value will survive for ages. This
means that the metal concerned was able to gain its specific
superconductive characteristics.

\par Now, we will calculate the energy the isotropic and anisotropic
distribution electrons exhibit. We will apply the normalization
condition formulated as follows: $$ G\hs\sum\limits_{\fbk}\hs
w_{\hh\fbk}=\overline N\hs  . \eqno (8.2) $$  Mean field
approximation electron energy takes on the following form: $$
\overline E=G\hs\sum\limits_{\fbk}\Bigl(\hs\varepsilon_{\fbk}\hs
w_{\hh\fbk} + \frac{1}{\hs 2\hs}\hs I\hs w_{\hh\fbk}\hs w_{\hh
-\hh\fbk} \Bigr)\hs . \eqno (8.3) $$

\par We will approximate dependence of electron kinetic energy
$\varepsilon_{\hh\fbk}$  from wave vector $\bf k$ by applying the
formula as follows: $$ \varepsilon_{\hh\fbk}=\frac{\hs\hbar^{\hh
2}\hh k^{\hh 2}\hh}{2\hh m}\hs , \eqno (8.4) $$ where $m$ is
effective itinerant electron mass. As provided by this formula, any
electron kinetic energy shall be counted from the band bottom to be
also called bottom of conduction band, i.e. $\varepsilon_{\hh\fbk\hh
=\hh 0}=0$.

\par To simplify calculations, instead of $\bf k$ summing we will
produce integration by $\varepsilon$ electron kinetic energy. By
applying the dependence (8.4) we will obtain the following symbolic
equation: $$ G\hs\sum_{\fbk}\hs ...\hs =A\hs\overline
N\hs\int\limits_0^\infty\hs ...\hs \hs\sqrt{\hs\varepsilon\hs}\hs\hs
d\varepsilon\hs , $$ where $$ A=\frac{\hs G\hs m\hs\sqrt{\hh 2\hh
m\hh}\hs V\hs} {2\hs\pi^{\hh 2}\hh\hbar^{\hh 3}\hh\overline N}\hs ,
\hskip 5mm \varepsilon=\frac{\hs\hbar^{\hh 2}\hh k^{\hh 2}\hh}{2\hh
m}\hs , \hskip 5mm d\varepsilon=\frac{\hs\hbar^{\hh 2}\hh k\hh dk
\hh}{m}\hs . $$ The upper integration limit may be equal to
$\infty$, since the occupational probability of states which energy
$\varepsilon$ is specified at the ceiling of conduction band is
actually equal to zero. Now, we will formulate the normalization
condition (8.2) by the method as follows: $$ A\int\limits_0^\infty
w(\varepsilon)\hs\sqrt{\hh\varepsilon\hs}\hs d\varepsilon=1\hs .
\eqno (8.5) $$ As for the isotropic distribution electron energy, we
will formulate the following formula: $$ \overline E^{\hh
(i)}=A\hs\overline N\hh\int\limits_0^\infty \biggl(\varepsilon
+\hs\frac{\hh 1\hh}{2}\hs I\hs w(\varepsilon)\biggr)\hs
w(\varepsilon)\hs\sqrt{\hh\varepsilon\hs}\hs d\varepsilon\hs . \eqno
(8.6) $$

\par If the isotropic electron distribution function applied at
$T=0$ is formulated according to (7.1), the equations (8.5) and
(8.6) take on the following form: $$ A\int\limits_0^{\mu
-I}\sqrt{\hh\varepsilon\hs}\hs d\varepsilon+
A\int\limits_{\mu-I}^\mu\frac{\hh\mu -\varepsilon\hh}{I}\hs
\sqrt{\hh\varepsilon\hs}\hs d\varepsilon=1\hs , $$ $$ \overline
E_{\rm o}^{\hh (i)}=A\hs\overline N\int\limits_0^{\mu-I}
\biggl(\varepsilon +\frac{\hh 1\hh}{2}\hs I\biggr)\hh
\sqrt{\hh\varepsilon\hs}\hs d\varepsilon\hs + \hs \frac{\hh
A\hs\overline N\hh}{2\hs I}\hh\int\limits_{\mu-I}^\mu
\bigl(\hh\mu^{\hh 2}-\varepsilon^{\hh 2}\hh\bigr)\hs
\sqrt{\hh\varepsilon\hs}\hs d\varepsilon\hs . $$ Since we apply the
small parameter $$ \lambda =\frac{I}{\hs\varepsilon_F}\hs , $$ where
$$ \varepsilon_F=\biggl(\frac{3}{\hs 2\hs A\hs}\biggr)^{2/3} $$ --
refers to Fermi energy, we will define that the chemical potential
and isotropic wave vector distribution electron energy at $T=0$ take
on the forms as follows: $$ \mu_{\rm o}
=\varepsilon_F\hh\biggl(1+\frac{\hh 1\hh}{2}\hs\lambda- \frac{1}{\hh
48\hh}\hs\lambda^{\hh 2}+...\biggr)\hs , $$ $$ \overline E_{\rm
o}^{\hh (i)}=\overline N\hs\varepsilon_F \hh\biggl(\frac{\hh
3\hh}{5}+\frac{\hh 1\hh}{2}\hs\lambda - \frac{1}{\hh
16\hh}\hs\lambda^{\hh 2}+...\biggr)\hs . \eqno (8.7) $$

\par We will assume that the anisotropic wave vector electron
distribution at $T=0$ is defined by the function as follows: $$
w_{\hh\fbk}=\left\{\begin{array}{ccl} 1 & \hbox{at} &
\varepsilon_{\hh\fbk}\leq\mu - I\hs , \medskip \\ 1 & \hbox{at} &
\mu -I<\varepsilon_{\hh\fbk}<\mu\hs , \hs\hs k_x>0 \medskip \\ 0 &
\hbox{at} & \mu -I<\varepsilon_{\hh\fbk}<\mu\hs , \hs\hs k_x<0
\medskip \\ 0 & \hbox{at} & \varepsilon_{\hh\fbk}\geq\mu\hs .
\end{array}\right. \eqno (8.8) $$ As provided by the above formula,
only one half of the $k_x>0$ states may be referred to as the
occupied ones to occur in layer $S$ above the Fermi surface, which
thickness $\delta k$ is proportional to interaction parameter $I$.
For anisotropic distribution pattern, see Fig. 8.

\unitlength=1mm \centerline{\begin{picture}(65,59)
\put(27,27){\circle*{0.7}}\put(12,12){\circle*{0.7}}
\put(42,42){\circle*{0.7}}\put(47,9){$\delta k$}
\put(27,0){\vector(0,1){54}}\put(22,52){$k_{\hh y}$}
\put(0,27){\vector(1,0){56}}\put(54,23){$k_{\hh x}$}
\put(27,27){\vector(1,1){14.6}}\put(32,35){${\bf k}$}
\put(27,27){\vector(-1,-1){14.6}}\put(14,20){${-\hh\bf k}$}
\put(2,8){$w_{\hh -\hh\fbk}=0$}\put(43,45){$w_{\hh\fbk}=1$}
\put(27,27){\vector(1,-1){12.5}}\put(48,6){\vector(-1,1){5}}
\put(27,27){\unitlength=1.2mm{\special{em:linewidth 0.3pt}}
\put(15,0){\special{em:moveto}} \put(14.94,1.31)
{\special{em:lineto}} \put(14.78,2.61) {\special{em:lineto}}
\put(14.49,3.89) {\special{em:lineto}} \put(14.10,5.13)
{\special{em:lineto}} \put(13.59,6.35) {\special{em:lineto}}
\put(12.99,7.5)  {\special{em:lineto}} \put(12.29,8.61)
{\special{em:lineto}} \put(11.49,9.65) {\special{em:lineto}}
\put(10.61,10.61){\special{em:lineto}} \put(9.65,11.49)
{\special{em:lineto}} \put(8.61,12.29) {\special{em:lineto}}
\put(7.5,12.99)  {\special{em:lineto}} \put(6.35,13.59)
{\special{em:lineto}} \put(5.13,14.10) {\special{em:lineto}}
\put(3.89,14.49) {\special{em:lineto}} \put(2.61,14.78)
{\special{em:lineto}} \put(1.31,14.94) {\special{em:lineto}}
\put(0,15)       {\special{em:lineto}} \put(0,15)
{\special{em:moveto}} \put(-1.31,14.94) {\special{em:lineto}}
\put(-2.61,14.78) {\special{em:lineto}} \put(-3.89,14.49)
{\special{em:lineto}} \put(-5.13,14.10) {\special{em:lineto}}
\put(-6.35,13.59) {\special{em:lineto}} \put(-7.5,12.99)
{\special{em:lineto}} \put(-8.61,12.29) {\special{em:lineto}}
\put(-9.65,11.49) {\special{em:lineto}}
\put(-10.61,10.61){\special{em:lineto}} \put(-11.49,9.65)
{\special{em:lineto}} \put(-12.29,8.61) {\special{em:lineto}}
\put(-12.99,7.5)  {\special{em:lineto}} \put(-13.59,6.35)
{\special{em:lineto}} \put(-14.10,5.13) {\special{em:lineto}}
\put(-14.49,3.89) {\special{em:lineto}} \put(-14.78,2.61)
{\special{em:lineto}} \put(-14.94,1.31) {\special{em:lineto}}
\put(-15,0)       {\special{em:lineto}} \put(-15,0)
{\special{em:moveto}} \put(-14.94,-1.31) {\special{em:lineto}}
\put(-14.78,-2.61) {\special{em:lineto}} \put(-14.49,-3.89)
{\special{em:lineto}} \put(-14.10,-5.13) {\special{em:lineto}}
\put(-13.59,-6.35) {\special{em:lineto}} \put(-12.99,-7.5)
{\special{em:lineto}} \put(-12.29,-8.61) {\special{em:lineto}}
\put(-11.49,-9.65) {\special{em:lineto}}
\put(-10.61,-10.61){\special{em:lineto}} \put(-9.65,-11.49)
{\special{em:lineto}} \put(-8.61,-12.29) {\special{em:lineto}}
\put(-7.5,-12.99)  {\special{em:lineto}} \put(-6.35,-13.59)
{\special{em:lineto}} \put(-5.13,-14.10) {\special{em:lineto}}
\put(-3.89,-14.49) {\special{em:lineto}} \put(-2.61,-14.78)
{\special{em:lineto}} \put(-1.31,-14.94) {\special{em:lineto}}
\put(0,-15) {\special{em:lineto}} \put(0,-15) {\special{em:moveto}}
\put(1.31,-14.94) {\special{em:lineto}} \put(2.61,-14.78)
{\special{em:lineto}} \put(3.89,-14.49) {\special{em:lineto}}
\put(5.13,-14.10) {\special{em:lineto}} \put(6.35,-13.59)
{\special{em:lineto}} \put(7.5,-12.99) {\special{em:lineto}}
\put(8.61,-12.29) {\special{em:lineto}} \put(9.65,-11.49)
{\special{em:lineto}} \put(10.61,-10.61){\special{em:lineto}}
\put(11.49,-9.65) {\special{em:lineto}} \put(12.29,-8.61)
{\special{em:lineto}} \put(12.99,-7.5)  {\special{em:lineto}}
\put(13.59,-6.35) {\special{em:lineto}} \put(14.10,-5.13)
{\special{em:lineto}} \put(14.49,-3.89) {\special{em:lineto}}
\put(14.78,-2.61) {\special{em:lineto}} \put(14.94,-1.31)
{\special{em:lineto}} \put(15,0)        {\special{em:lineto}} }
\put(27,27){\unitlength=1.5mm{\special{em:linewidth 0.3pt}}
\put(15,0){\special{em:moveto}} \put(14.94,1.31)
{\special{em:lineto}} \put(14.78,2.61) {\special{em:lineto}}
\put(14.49,3.89) {\special{em:lineto}} \put(14.10,5.13)
{\special{em:lineto}} \put(13.59,6.35) {\special{em:lineto}}
\put(12.99,7.5)  {\special{em:lineto}} \put(12.29,8.61)
{\special{em:lineto}} \put(11.49,9.65) {\special{em:lineto}}
\put(10.61,10.61){\special{em:lineto}} \put(9.65,11.49)
{\special{em:lineto}} \put(8.61,12.29) {\special{em:lineto}}
\put(7.5,12.99)  {\special{em:lineto}} \put(6.35,13.59)
{\special{em:lineto}} \put(5.13,14.10) {\special{em:lineto}}
\put(3.89,14.49) {\special{em:lineto}} \put(2.61,14.78)
{\special{em:lineto}} \put(1.31,14.94) {\special{em:lineto}}
\put(0,15) {\special{em:lineto}} \put(0,-15) {\special{em:moveto}}
\put(1.31,-14.94) {\special{em:lineto}} \put(2.61,-14.78)
{\special{em:lineto}} \put(3.89,-14.49) {\special{em:lineto}}
\put(5.13,-14.10) {\special{em:lineto}} \put(6.35,-13.59)
{\special{em:lineto}} \put(7.5,-12.99) {\special{em:lineto}}
\put(8.61,-12.29) {\special{em:lineto}} \put(9.65,-11.49)
{\special{em:lineto}} \put(10.61,-10.61){\special{em:lineto}}
\put(11.49,-9.65) {\special{em:lineto}} \put(12.29,-8.61)
{\special{em:lineto}} \put(12.99,-7.5)  {\special{em:lineto}}
\put(13.59,-6.35) {\special{em:lineto}} \put(14.10,-5.13)
{\special{em:lineto}} \put(14.49,-3.89) {\special{em:lineto}}
\put(14.78,-2.61) {\special{em:lineto}} \put(14.94,-1.31)
{\special{em:lineto}} \put(15,0)        {\special{em:lineto}} }
\end{picture}}

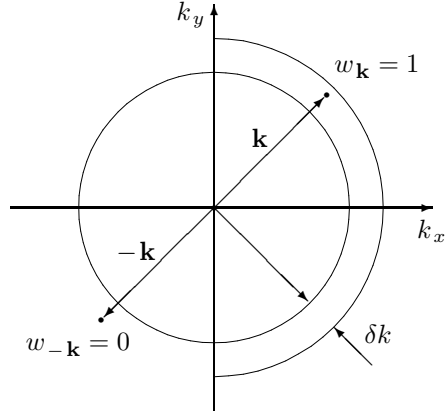
\vskip 3mm \centerline{\it Fig. 8. Anisotropic distribution function
at $T=0$.} \vskip 3mm

\par Here, the normalization condition gives rise to the following
equation: $$ A\int\limits_0^{\mu -I}\sqrt{\hh\varepsilon\hs}\hs
d\varepsilon+ \frac{1}{\hh 2\hh}\hs A\int\limits_{\mu-I}^\mu
\sqrt{\hh\varepsilon\hs}\hs d\varepsilon=1\hs . $$ As for electron
energy, it may be calculated by the formula as follows: $$ \overline
E_{\rm o}^{\hh (s)}=A\hs\overline N\int\limits_0^{\mu-I}
\biggl(\varepsilon +\frac{\hh 1\hh}{2}\hs I\biggr)\hh
\sqrt{\hh\varepsilon\hs}\hs d\varepsilon + \frac{\hh 1\hh}{2}\hs
A\hs\overline N\hs\int\limits_{\mu-I}^\mu
\varepsilon\hs\sqrt{\hh\varepsilon\hs}\hs d\varepsilon\hs . $$ As
provided by the above calculation, the following formulation is
obtained: $$ \mu =\varepsilon_F\hh\biggl(1+\frac{\hh
1\hh}{2}\hs\lambda- \frac{1}{\hh 16\hh}\hs\lambda^{\hh
2}+...\biggr)\hs , $$ $$ \overline E_{\rm o}^{\hh (s)}=\overline
N\hs\varepsilon_F \hh\biggl(\frac{\hh 3\hh}{5}+\frac{\hh
1\hh}{2}\hs\lambda- \frac{3}{\hh 16\hh}\hs\lambda^{\hh
2}+...\biggr)\hs . \eqno (8.9) $$ If occupied and free state pairs
that match specific wave vectors $\bf k$ and $-\hh\bf k$ will be
distributed within layer $S$ by any other way, the chemical
potential and electron energy rating will remain the same.

\par The difference of electron energy values (8.7) and (8.9) will
be formulated by the equation as follows: $$ \overline E_{\rm
o}^{\hh (i)}-\overline E_{\rm o}^{\hh (s)}= \frac{\hs\overline N\hs
I^{\hh 2}\hh}{\hh 8 \hs\varepsilon_F\hs}>0\hs . $$ Thus, we get to
the conclusion that the state of itinerant electrons described by
the anisotropic distribution function is the primary one - i.e. the
electron system specified in this condition is of the lowest
energy.

\par Considering for the aforesaid about the anisotropic electron
energy distribution we will plot the pattern of superconductive
state, as shown in Fig. 9.

\vskip 7mm\centerline{\unitlength=1.2mm\begin{picture}(87,58)
\put(0,10){\vector(1,0){87}}\put(79,5){$\varepsilon-\mu$}
\multiput(20,10.5)(0,2){20}{\line(0,1){1.4}}
\multiput(0,50.5)(2,0){31}{\line(1,0){1.4}}
\put(20,10){\line(0,-1){1}}\put(17,5){$-I$}
\put(40,10){\line(0,-1){1}}\put(36.9,5){$-\frac12\hh I$}
\put(60,10){\line(0,-1){1}}\put(59.2,5){0}
\put(60,10){\vector(0,1){50}}\put(62,58.5){$w$} 
\put(62,49){1,0} \put(60,30){\line(1,0){1}}\put(62,29){0,5}
\put(60,10){\unitlength=1.2mm\special{em:linewidth 0.3pt}
\put(27.00,1.2) {\special{em:moveto}}
\put(25.99,1.3){\special{em:lineto}} \put(21.55,2)
{\special{em:lineto}} \put(17.10,3) {\special{em:lineto}}
\put(13.58,4)  {\special{em:lineto}} \put(10.57,5)
{\special{em:lineto}} \put(7.880,6) {\special{em:lineto}}
\put(5.408,7)  {\special{em:lineto}} \put(3.088,8)
{\special{em:lineto}} \put(0.8960,9) {\special{em:lineto}}
\put(-1.208,10)  {\special{em:lineto}} \put(-1.208,10)
{\special{em:lineto}} \put(-3.244,11) {\special{em:lineto}} }
\put(60,10){\unitlength=1.2mm\special{em:linewidth 0.3pt}
\put(-37,29) {\special{em:moveto}} \put(-36.76,29)
{\special{em:lineto}} \put(-38.79,30) {\special{em:lineto}}
\put(-40.88,31)  {\special{em:lineto}} \put(-43.08,32)
{\special{em:lineto}} \put(-45.40,33) {\special{em:lineto}}
\put(-47.88,34)  {\special{em:lineto}} \put(-50.56,35)
{\special{em:lineto}} \put(-53.56,36) {\special{em:lineto}}
\put(-57.08,37)  {\special{em:lineto}} \put(-60.00,37.7)
{\special{em:lineto}} }
\put(60,10){\unitlength=1.2mm\special{em:linewidth 0.3pt}
\put(-36.61,28.25)  {\special{em:moveto}} \put(-36.56,27.54)
{\special{em:lineto}} \put(-36.49,26.81)  {\special{em:lineto}}
\put(-36.30,26.00)  {\special{em:lineto}} \put(-36.16,25.22)
{\special{em:lineto}} \put(-35.90,24.36)  {\special{em:lineto}}
\put(-35.60,23.46)  {\special{em:lineto}} \put(-35.30,22.58)
{\special{em:lineto}} \put(-34.92,21.62)  {\special{em:lineto}}
\put(-34.49,20.64)  {\special{em:lineto}} \put(-34.04,19.66)
{\special{em:lineto}} \put(-33.52,18.62)  {\special{em:lineto}}
\put(-32.93,17.55)  {\special{em:lineto}} \put(-32.27,16.47)
{\special{em:lineto}} \put(-31.54,15.34)  {\special{em:lineto}}
\put(-30.70,14.16)  {\special{em:lineto}} \put(-29.75,12.97)
{\special{em:lineto}} \put(-28.64,11.70)  {\special{em:lineto}}
\put(-27.28,10.36)  {\special{em:lineto}} \put(-25.57,8.904)
{\special{em:lineto}} \put(-23.82,7.696)  {\special{em:lineto}}
\put(-22.74,7.064)  {\special{em:lineto}} \put(-20.56,6.016)
{\special{em:lineto}} \put(-17.259,4.936)  {\special{em:lineto}}
\put(-13.507,4.348)  {\special{em:lineto}} \put(-11.362,4.300)
{\special{em:lineto}} \put(-9.7588,4.428)  {\special{em:lineto}}
\put(-8.4684,4.664)  {\special{em:lineto}} \put(-7.3924,4.988)
{\special{em:lineto}} \put(-6.4840,5.376)  {\special{em:lineto}}
\put(-5.7180,5.848)  {\special{em:lineto}} \put(-5.0736,6.376)
{\special{em:lineto}} \put(-4.5384,6.984)  {\special{em:lineto}}
\put(-4.1092,7.640)  {\special{em:lineto}} \put(-3.7788,8.388)
{\special{em:lineto}} \put(-3.5472,9.188)  {\special{em:lineto}}
\put(-3.4080,10.12)  {\special{em:lineto}} \put(-3.379, 11.74)
{\special{em:lineto}} \put(-3.4399,12.46)  {\special{em:lineto}}
\put(-3.5434,13.20)  {\special{em:lineto}} \put(-3.6889,13.99)
{\special{em:lineto}} \put(-3.8770,14.80)  {\special{em:lineto}}
\put(-4.1085,15.65)  {\special{em:lineto}} \put(-4.3842,16.53)
{\special{em:lineto}} \put(-4.7046,17.43)  {\special{em:lineto}}
\put(-5.0727,18.38)  {\special{em:lineto}} \put(-5.4901,19.35)
{\special{em:lineto}} \put(-5.9590,20.35)  {\special{em:lineto}}
\put(-6.4850,21.38)  {\special{em:lineto}} \put(-7.0731,22.44)
{\special{em:lineto}} \put(-7.7300,23.54)  {\special{em:lineto}}
\put(-8.4674,24.67)  {\special{em:lineto}} \put(-9.3005,25.83)
{\special{em:lineto}} \put(-10.252,27.04)  {\special{em:lineto}}
\put(-11.364,28.30)  {\special{em:lineto}} \put(-12.705,29.63)
{\special{em:lineto}} \put(-14.444,31.10)  {\special{em:lineto}}
\put(-15.347,31.76)  {\special{em:lineto}} \put(-16.176,32.30)
{\special{em:lineto}} \put(-17.259,32.94)  {\special{em:lineto}}
\put(-19.453,33.99)  {\special{em:lineto}} \put(-22.74,35.06)
{\special{em:lineto}} \put(-26.49,35.65)  {\special{em:lineto}}
\put(-28.64,35.70)  {\special{em:lineto}} \put(-30.24,35.57)
{\special{em:lineto}} \put(-31.54,35.34)  {\special{em:lineto}}
\put(-32.61,35.01)  {\special{em:lineto}} \put(-33.52,34.62)
{\special{em:lineto}} \put(-34.28,34.15)  {\special{em:lineto}}
\put(-34.92,33.62)  {\special{em:lineto}} \put(-35.45,33.02)
{\special{em:lineto}} \put(-35.90,32.36)  {\special{em:lineto}}
\put(-36.23,31.61)  {\special{em:lineto}} \put(-36.49,30.81)
{\special{em:lineto}} \put(-36.55,29.88)  {\special{em:lineto}}
\put(-36.61,28.25)  {\special{em:lineto}} } \end{picture}}

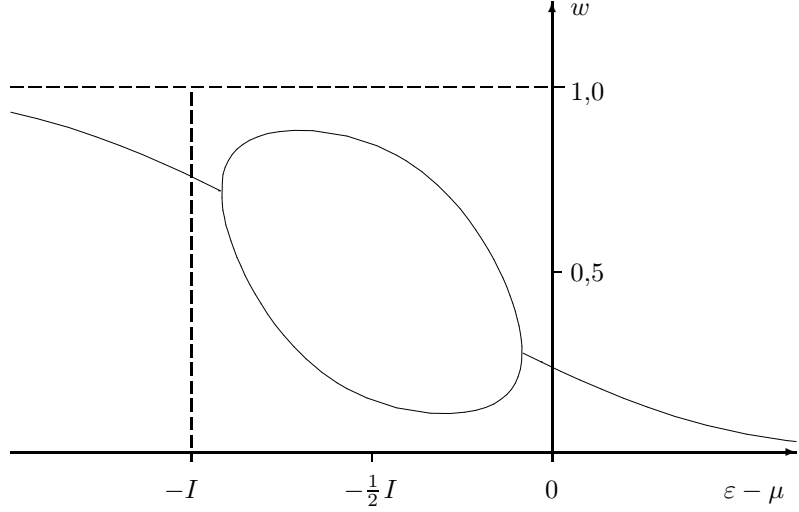
\vskip -3mm\centerline {\it Fig. 9. Anisotropic distribution of
energy $\varepsilon$ conductivity electrons } \centerline {\it
subject to the lowest energy $\overline E$ at the temperature of
$\tau=0.8$. }

\vskip 5mm \centerline{\bf 9. Maximum superconductivity electron
velocity at $T=0$} \vskip 2mm

\par Now, we will find ordered electron motion velocity in the state
described by the distribution function (7.2) at the temperature of
$T=0$. For this purpose, we well substitute the wave vector sum
specified in the formula (8.1) for the following integral: $$ {\bf
v}=\frac{\hs G\hs\hbar\hs V\hs} {\hs 8\hs\pi^{\hh 3}\hs
m\hs\overline N\hs}\hs \int{\bf k}\hs w_{\hh\fbk}\hs d^{\hh 3}k\hs .
\eqno (9.1) $$ If $T=0$, the inner and outer radii of layer $S$
shall be respectively equal to as follows: $$
k_1=\frac{1}{\hs\hbar\hs}\hs\sqrt{\hs 2\hs m\hs (\hh\mu
-I\hh)\hs}\hs , \hskip 10mm k_2=\frac{1}{\hs\hbar\hs}\hs\sqrt{\hs
2\hs m\hs\mu\hs}\hs . $$ In this case, the layer thickness is
defined by the following formula: $$ \delta
k=k_2-k_1=\frac{I}{\hs\hbar\hs}\hs \sqrt{\hs\frac{\hs m\hs}{\hs
2\hs\varepsilon_F\hh }\hh}\hs . \eqno (9.2) $$ If the states are
filled with electrons in one half of layer $S$, but another one
remains free, the rate of ordered electron motion velocity exhibits
its maximum value $$ v_{max}=\frac{\hs 3\hs I\hs} {\hs 4\hs\sqrt{\hs
2\hs m\hs\varepsilon_F}\hs}\hs . \eqno (9.3) $$

\vskip 5mm \centerline{\bf 10. Electric current-forced
superconducting state destroy effect } \vskip 2mm

\par Superconducting state of the itinerant electrons is destroyed
in the events when the current running over metal items exceeds its
particular critical value (Silsbee effect). We will assume that
specific homogeneous electric field with tension $\bf E$ is produced
inside the metal. Under the effect of this field the itinerant
electrons will execute their ordered motion at the average velocity
$\bf u$ which direction agrees with another one that affecting force
electron $-\hh e\hh\bf E$. In this case, electron state distribution
function $w_{\hh\fbk}$ may be defined by the equation (3.3), where
electron energy $\overline\varepsilon_{\fbk}$ depends on wave vector
$\bf k$ in the following manner: $$
\overline\varepsilon_{\fbk}=\varepsilon_{\fbk}-\hbar\hs{\bf
k}\hs{\bf u}+ I\hs w_{\hh -\hh\fbk}\hs . \eqno (10.1) $$ It is no
easy matter to make exact solution of the equation (3.3). Therefore,
we will consider its approximate solution only. We will assume that
distribution function $w_{\hh\fbk}$ has the form as follows: $$
w_{\hh\fbk}=f({\bf k}-{\bf k}_{\hh\rm o})\hs , \eqno (10.2) $$ where
$w_{\hh\fbk}=f({\bf k})$ is the solution of the equation (3.3),
providing that ${\bf u}=0$; $$ {\bf k}_{\hh\rm o}=\frac{\hs m\hs{\bf
u}\hs}{\hbar}\hs . $$ Function $w_{\hh\fbk}=f({\bf k})$ is equal to
a unity almost at all the points occurred inside Fermi sphere:
$k<k_F$, excluding the points at surface $S$ of that sphere. If
occurred outside Fermi sphere, function $w_{\hh\fbk}=f({\bf k})$
almost everywhere is equal to zero. Field $D$ containing nonzero
function (10.2) is limited by the sphere with radius $k_F$, which
center $C$ is displaced off the origin of coordinates $O$ by vector
${\bf k}_{\hh\rm o}$. In other words, the wave vectors that agree
with occupied electronic states belong to field $D$. The theory
under discussion is applied to superconductivity to be due to
interaction of electrons, which wave vectors $\bf k$ and $-\hh\bf k$
belong to spherical layer $S$ with its radius equal to $k_F$ and
thickness - to $\delta k$. No anisotropy is created with the
electrons distributed over their wave vectors, when the rate of
displacement $k_{\hh\rm o}$ of field $D$ is so large that one half
of layer $S$ finds itself beyond this field. For arrangement of
field $D$ and layer $S$, see Fig. 10.

\unitlength=1.2mm \centerline{\begin{picture}(65,50)
\put(27,20){\circle*{0.7}}\put(23.5,19.5){$O$}
\put(27,36.5){\circle*{0.7}}\put(24,32.8){$A$}
\put(34,38){\circle*{0.7}}\put(32.5,39){$B$}
\multiput(34,20)(0,1.85){10}{\line(0,1){1.4}}
\put(34,20){\circle*{0.7}}\put(32,16.5){$C$}
\multiput(34,20)(-0.82,1.9){9}{\unitlength=1.2mm{\special{em:linewidth
0.3pt}}
\put(0,0){\special{em:moveto}}\put(-0.63,1.4){\special{em:lineto}} }
\put(27,20){\thicklines\vector(1,0){7}}\put(28.5,21.5){${\bf
k}_{\hh\rm o}$} \put(27,20){\vector(-1,-1){19}}\put(4,1){$k_x$}
\put(27,20){\vector(1,0){35}}\put(60,16){$k_y$}
\put(27,20){\vector(0,1){27}}\put(23,45){$k_z$}
\put(34,20){\vector(2,-1){16}}\put(46.5,14.5){$k_F$}
\put(2,20){\vector(1,0){7}}\put(3.5,21){$\delta k$}
\put(14.5,20){\vector(-1,0){4}} \put(11,33){$S$}\put(48,33){$D$}
 \put(27,20){\unitlength=1.3mm{\special{em:linewidth 0.3pt}}
 \put(15,0)       {\special{em:moveto}} \put(14.94,1.31)
{\special{em:lineto}} \put(14.78,2.61) {\special{em:lineto}}
\put(14.49,3.89) {\special{em:lineto}} \put(14.10,5.13)
{\special{em:lineto}} \put(13.59,6.35) {\special{em:lineto}}
\put(12.99,7.5)  {\special{em:lineto}} \put(12.29,8.61)
{\special{em:lineto}} \put(11.49,9.65) {\special{em:lineto}}
\put(10.61,10.61){\special{em:lineto}} \put(9.65,11.49)
{\special{em:lineto}} \put(8.61,12.29) {\special{em:lineto}}
\put(7.5,12.99)  {\special{em:lineto}} \put(6.35,13.59)
{\special{em:lineto}} \put(5.13,14.10) {\special{em:lineto}}
\put(3.89,14.49) {\special{em:lineto}} \put(2.61,14.78)
{\special{em:lineto}} \put(1.31,14.94) {\special{em:lineto}}
\put(0,15)       {\special{em:lineto}} \ \put(0,15)
{\special{em:moveto}} \put(-1.31,14.94) {\special{em:lineto}}
\put(-2.61,14.78) {\special{em:lineto}} \put(-3.89,14.49)
{\special{em:lineto}} \put(-5.13,14.10) {\special{em:lineto}}
\put(-6.35,13.59) {\special{em:lineto}} \put(-7.5,12.99)
{\special{em:lineto}} \put(-8.61,12.29) {\special{em:lineto}}
\put(-9.65,11.49) {\special{em:lineto}}
\put(-10.61,10.61){\special{em:lineto}} \put(-11.49,9.65)
{\special{em:lineto}} \put(-12.29,8.61) {\special{em:lineto}}
\put(-12.99,7.5)  {\special{em:lineto}} \put(-13.59,6.35)
{\special{em:lineto}} \put(-14.10,5.13) {\special{em:lineto}}
\put(-14.49,3.89) {\special{em:lineto}} \put(-14.78,2.61)
{\special{em:lineto}} \put(-14.94,1.31) {\special{em:lineto}}
\put(-15,0)       {\special{em:lineto}} \put(-15,0)
{\special{em:moveto}} \put(-14.94,-1.31) {\special{em:lineto}}
\put(-14.78,-2.61) {\special{em:lineto}} \put(-14.49,-3.89)
{\special{em:lineto}} \put(-14.10,-5.13) {\special{em:lineto}}
\put(-13.59,-6.35) {\special{em:lineto}} \put(-12.99,-7.5)
{\special{em:lineto}} \put(-12.29,-8.61) {\special{em:lineto}}
\put(-11.49,-9.65) {\special{em:lineto}}
\put(-10.61,-10.61){\special{em:lineto}} \put(-9.65,-11.49)
{\special{em:lineto}} \put(-8.61,-12.29) {\special{em:lineto}}
\put(-7.5,-12.99)  {\special{em:lineto}} \put(-6.35,-13.59)
{\special{em:lineto}} \put(-5.13,-14.10) {\special{em:lineto}}
\put(-3.89,-14.49) {\special{em:lineto}} \put(-2.61,-14.78)
{\special{em:lineto}} \put(-1.31,-14.94) {\special{em:lineto}}
\put(0,-15)        {\special{em:lineto}}  \put(0,-15)
{\special{em:moveto}} \put(1.31,-14.94) {\special{em:lineto}}
\put(2.61,-14.78) {\special{em:lineto}} \put(3.89,-14.49)
{\special{em:lineto}} \put(5.13,-14.10) {\special{em:lineto}}
\put(6.35,-13.59) {\special{em:lineto}} \put(7.5,-12.99)
{\special{em:lineto}} \put(8.61,-12.29) {\special{em:lineto}}
\put(9.65,-11.49) {\special{em:lineto}}
\put(10.61,-10.61){\special{em:lineto}} \put(11.49,-9.65)
{\special{em:lineto}} \put(12.29,-8.61) {\special{em:lineto}}
\put(12.99,-7.5)  {\special{em:lineto}} \put(13.59,-6.35)
{\special{em:lineto}} \put(14.10,-5.13) {\special{em:lineto}}
\put(14.49,-3.89) {\special{em:lineto}} \put(14.78,-2.61)
{\special{em:lineto}} \put(14.94,-1.31) {\special{em:lineto}}
\put(15,0)        {\special{em:lineto}} }
\put(26.9,20){\unitlength=1.44mm{\special{em:linewidth 0.3pt}}
 \put(15,0)       {\special{em:moveto}} \put(14.94,1.31)
{\special{em:lineto}} \put(14.78,2.61) {\special{em:lineto}}
\put(14.49,3.89) {\special{em:lineto}} \put(14.10,5.13)
{\special{em:lineto}} \put(13.59,6.35) {\special{em:lineto}}
\put(12.99,7.5)  {\special{em:lineto}} \put(12.29,8.61)
{\special{em:lineto}} \put(11.49,9.65) {\special{em:lineto}}
\put(10.61,10.61){\special{em:lineto}} \put(9.65,11.49)
{\special{em:lineto}} \put(8.61,12.29) {\special{em:lineto}}
\put(7.5,12.99)  {\special{em:lineto}} \put(6.35,13.59)
{\special{em:lineto}} \put(5.13,14.10) {\special{em:lineto}}
\put(3.89,14.49) {\special{em:lineto}} \put(2.61,14.78)
{\special{em:lineto}} \put(1.31,14.94) {\special{em:lineto}}
\put(0,15)       {\special{em:lineto}}  \put(0,15)
{\special{em:moveto}} \put(-1.31,14.94) {\special{em:lineto}}
\put(-2.61,14.78) {\special{em:lineto}} \put(-3.89,14.49)
{\special{em:lineto}} \put(-5.13,14.10) {\special{em:lineto}}
\put(-6.35,13.59) {\special{em:lineto}} \put(-7.5,12.99)
{\special{em:lineto}} \put(-8.61,12.29) {\special{em:lineto}}
\put(-9.65,11.49) {\special{em:lineto}}
\put(-10.61,10.61){\special{em:lineto}} \put(-11.49,9.65)
{\special{em:lineto}} \put(-12.29,8.61) {\special{em:lineto}}
\put(-12.99,7.5)  {\special{em:lineto}} \put(-13.59,6.35)
{\special{em:lineto}} \put(-14.10,5.13) {\special{em:lineto}}
\put(-14.49,3.89) {\special{em:lineto}} \put(-14.78,2.61)
{\special{em:lineto}} \put(-14.94,1.31) {\special{em:lineto}}
\put(-15,0)       {\special{em:lineto}} \put(-15,0)
{\special{em:moveto}} \put(-14.94,-1.31) {\special{em:lineto}}
\put(-14.78,-2.61) {\special{em:lineto}} \put(-14.49,-3.89)
{\special{em:lineto}} \put(-14.10,-5.13) {\special{em:lineto}}
\put(-13.59,-6.35) {\special{em:lineto}} \put(-12.99,-7.5)
{\special{em:lineto}} \put(-12.29,-8.61) {\special{em:lineto}}
\put(-11.49,-9.65) {\special{em:lineto}}
\put(-10.61,-10.61){\special{em:lineto}} \put(-9.65,-11.49)
{\special{em:lineto}} \put(-8.61,-12.29) {\special{em:lineto}}
\put(-7.5,-12.99)  {\special{em:lineto}} \put(-6.35,-13.59)
{\special{em:lineto}} \put(-5.13,-14.10) {\special{em:lineto}}
\put(-3.89,-14.49) {\special{em:lineto}} \put(-2.61,-14.78)
{\special{em:lineto}} \put(-1.31,-14.94) {\special{em:lineto}}
\put(0,-15)        {\special{em:lineto}}  \put(0,-15)
{\special{em:moveto}} \put(1.31,-14.94) {\special{em:lineto}}
\put(2.61,-14.78) {\special{em:lineto}} \put(3.89,-14.49)
{\special{em:lineto}} \put(5.13,-14.10) {\special{em:lineto}}
\put(6.35,-13.59) {\special{em:lineto}} \put(7.5,-12.99)
{\special{em:lineto}} \put(8.61,-12.29) {\special{em:lineto}}
\put(9.65,-11.49) {\special{em:lineto}}
\put(10.61,-10.61){\special{em:lineto}} \put(11.49,-9.65)
{\special{em:lineto}} \put(12.29,-8.61) {\special{em:lineto}}
\put(12.99,-7.5)  {\special{em:lineto}} \put(13.59,-6.35)
{\special{em:lineto}} \put(14.10,-5.13) {\special{em:lineto}}
\put(14.49,-3.89) {\special{em:lineto}} \put(14.78,-2.61)
{\special{em:lineto}} \put(14.94,-1.31) {\special{em:lineto}}
\put(15,0)        {\special{em:lineto}} }
\put(34,20){\unitlength=1.44mm{\special{em:linewidth 0.3pt}}
\put(15,0)       {\special{em:moveto}} \put(14.94,1.31)
{\special{em:lineto}} \put(14.78,2.61) {\special{em:lineto}}
\put(14.49,3.89) {\special{em:lineto}} \put(14.10,5.13)
{\special{em:lineto}} \put(13.59,6.35) {\special{em:lineto}}
\put(12.99,7.5)  {\special{em:lineto}} \put(12.29,8.61)
{\special{em:lineto}} \put(11.49,9.65) {\special{em:lineto}}
\put(10.61,10.61){\special{em:lineto}} \put(9.65,11.49)
{\special{em:lineto}} \put(8.61,12.29) {\special{em:lineto}}
\put(7.5,12.99)  {\special{em:lineto}} \put(6.35,13.59)
{\special{em:lineto}} \put(5.13,14.10) {\special{em:lineto}}
\put(3.89,14.49) {\special{em:lineto}} \put(2.61,14.78)
{\special{em:lineto}} \put(1.31,14.94) {\special{em:lineto}}
\put(0,15)       {\special{em:lineto}}  \put(0,15)
{\special{em:moveto}} \put(-1.31,14.94) {\special{em:lineto}}
\put(-2.61,14.78) {\special{em:lineto}} \put(-3.89,14.49)
{\special{em:lineto}} \put(-5.13,14.10) {\special{em:lineto}}
\put(-6.35,13.59) {\special{em:lineto}} \put(-7.5,12.99)
{\special{em:lineto}} \put(-8.61,12.29) {\special{em:lineto}}
\put(-9.65,11.49) {\special{em:lineto}}
\put(-10.61,10.61){\special{em:lineto}} \put(-11.49,9.65)
{\special{em:lineto}} \put(-12.29,8.61) {\special{em:lineto}}
\put(-12.99,7.5)  {\special{em:lineto}} \put(-13.59,6.35)
{\special{em:lineto}} \put(-14.10,5.13) {\special{em:lineto}}
\put(-14.49,3.89) {\special{em:lineto}} \put(-14.78,2.61)
{\special{em:lineto}} \put(-14.94,1.31) {\special{em:lineto}}
\put(-15,0)       {\special{em:lineto}}  \put(-15,0)
{\special{em:moveto}} \put(-14.94,-1.31) {\special{em:lineto}}
\put(-14.78,-2.61) {\special{em:lineto}} \put(-14.49,-3.89)
{\special{em:lineto}} \put(-14.10,-5.13) {\special{em:lineto}}
\put(-13.59,-6.35) {\special{em:lineto}} \put(-12.99,-7.5)
{\special{em:lineto}} \put(-12.29,-8.61) {\special{em:lineto}}
\put(-11.49,-9.65) {\special{em:lineto}}
\put(-10.61,-10.61){\special{em:lineto}} \put(-9.65,-11.49)
{\special{em:lineto}} \put(-8.61,-12.29) {\special{em:lineto}}
\put(-7.5,-12.99)  {\special{em:lineto}} \put(-6.35,-13.59)
{\special{em:lineto}} \put(-5.13,-14.10) {\special{em:lineto}}
\put(-3.89,-14.49) {\special{em:lineto}} \put(-2.61,-14.78)
{\special{em:lineto}} \put(-1.31,-14.94) {\special{em:lineto}}
\put(0,-15)        {\special{em:lineto}} \put(0,-15)
{\special{em:moveto}} \put(1.31,-14.94) {\special{em:lineto}}
\put(2.61,-14.78) {\special{em:lineto}} \put(3.89,-14.49)
{\special{em:lineto}} \put(5.13,-14.10) {\special{em:lineto}}
\put(6.35,-13.59) {\special{em:lineto}} \put(7.5,-12.99)
{\special{em:lineto}} \put(8.61,-12.29) {\special{em:lineto}}
\put(9.65,-11.49) {\special{em:lineto}}
\put(10.61,-10.61){\special{em:lineto}} \put(11.49,-9.65)
{\special{em:lineto}} \put(12.29,-8.61) {\special{em:lineto}}
\put(12.99,-7.5)  {\special{em:lineto}} \put(13.59,-6.35)
{\special{em:lineto}} \put(14.10,-5.13) {\special{em:lineto}}
\put(14.49,-3.89) {\special{em:lineto}} \put(14.78,-2.61)
{\special{em:lineto}} \put(14.94,-1.31) {\special{em:lineto}}
\put(15,0)        {\special{em:lineto}} } \end{picture}}

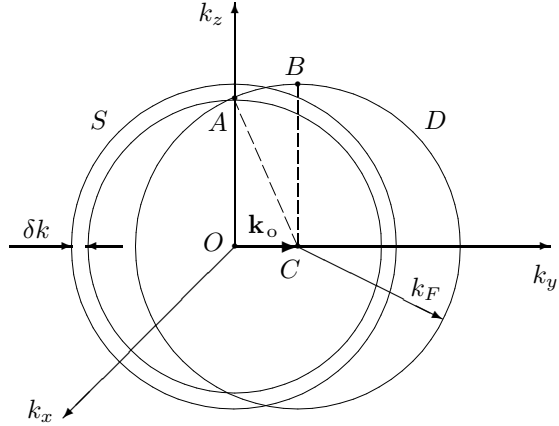
\vskip 3mm\centerline{\it Fig. 10. Displacement of Fermi sphere
under influence of electric field. } \vskip 3mm

\par Point $A$ belongs simultaneously to the displaced Fermi sphere
and to the inner surface of layer $S$. Therefore, $$ AC=k_F\hs ,
\hskip 7mm AO=k_F-\delta k\hs . $$ We will apply the rule of
Pythagoras for $AOC$ triangle. Now, we will gain the following
formula: $$ k_F^{\hh 2}=k_{\hh\rm o}^{\hh 2}+(\hh k_F-\delta k\hh
)^2\hs . $$ It appears from this equation that
superconductivity-force displacement $k_{\hh\rm o}$ will be
formulated as follows: $$ k_{\hh\rm o}=\sqrt{\hs 2\hs k_F\hs\delta
k\hs}\hs . $$ This formula may be transformed to: $$ k_{\hh\rm
o}=\frac{\hs\sqrt{\hs 2\hs m\hs\delta\varepsilon\hs}\hs}{\hbar}\hs ,
$$ where $\delta\varepsilon$ is energy width of the layer $S$. In
this connection, the average current velocity will be calculated by
the following equation: $$ u=\sqrt{\frac{\hs
2\hs\delta\varepsilon}{m}\hh}\hs . \eqno (10.3) $$ By this means
that the superconductive state of conduction electrons will be
destroyed when the external electric field makes them moving to the
same direction and produces the current, which density exceeds the
following value: $$ j_{\hbox{\it кр}}=e\hs n\hs u\hs . $$

\par We will specify the relation of maximum superconductive current
velocity $v_{max}$ to the least current speed $u$ destroying the
superconductive state: $$ \frac{\hs v_{max}\hh}{u}=\frac{\hs
3\hs}{8}\hs\sqrt{\hh\frac{\hs\delta\varepsilon\hs}{\hs\varepsilon_F\hh}\hh}
\hs . \eqno (10.4) $$ It is clear enough that the above relation is
much less than unity $v_{max}\ll u$.

\par  On cutting off the electric field and after thermodynamic
equilibration the electrons shall transform to their superconductive
state distributing as provided by the formula (7.2). Thereafter, the
average electron velocity shall drop down to value $v_{max}$ and
superconductive current of maximum density $j_{max}$ will run over
the metal.

\vskip 5mm \centerline{\bf 11. Mean energy dependence of kinetic
energy } \vskip 2mm

\par The rate of mean electron $\overline\varepsilon_{\fbk}$ energy
dependence of its kinetic energy $\varepsilon_{\fbk}$ is defined by
formula (4.2). As may be inferred from the above formula, the
electron energy with wave vector $\bf k$ depends on whether the
$-\hh\bf k$ wave vector state is free or occupied. Electron energy
$\overline\varepsilon$ may be specified by the $\varepsilon$ kinetic
energy functional form as follows: $$
\overline\varepsilon(\varepsilon)=\varepsilon + I\hs
w_1(\varepsilon)\hs . $$ or the pattern of this function at various
temperatures, see Fig. 11.

\unitlength=1.2mm\centerline{\begin{picture}(85,65)\put(23,2){
1}\put(27,11){\it 2}\put(30,19){\it 3}\put(7,25){\it 1}
\put(0,35){\vector(1,0){85}}\put(78,31.7){$\varepsilon$\hs--\hs$\mu$}
\put(61,32){0} \multiput(20,-5)(0,2.1){20}{\line(0,1){1.4}}
\put(20,35){\line(0,1){1}}\put(14,36){$-\hh I$}
\put(40,35){\line(0,1){1}}\put(38,38){$-\hh\frac{I}{\hh 2\hh}$}
\put(60,15){\line(1,0){1}}\put(62,14){$-\hh\frac{I}{\hh 2\hh}$}
\put(60,-10){\vector(0,1){71}}\put(62,58){$\overline\varepsilon$\hs--\hs$\mu$}
\multiput(20,35)(1.4,1.4){18}{\unitlength=1.2mm\special{em:linewidth
0.3pt} \put(0,0){\special{em:moveto}}\put(1,1){\special{em:lineto}}
}\put(0,15){\line(1,1){20}}\put(20,-5){\line(1,1){61}}
\multiput(20,-5)(2.1,0){20}{\line(1,0){1.4}}\put(62,-6){$-\hh{I}$}
\put(60,35){\unitlength=1.2mm\special{em:linewidth 0.3pt}
\put(21.000,23.20){\special{em:moveto}}
\put(19.165,21.66){\special{em:lineto}}
\put(17.098,20.10){\special{em:lineto}}
\put(15.257,18.76){\special{em:lineto}}
\put(13.578,17.58){\special{em:lineto}}
\put(12.024,16.53){\special{em:lineto}}
\put(10.567,15.57){\special{em:lineto}}
\put(9.1900,14.69){\special{em:lineto}}
\put(7.8770,13.88){\special{em:lineto}}
\put(6.6180,13.12){\special{em:lineto}}
\put(5.4050,12.41){\special{em:lineto}}
\put(4.2300,11.73){\special{em:lineto}}
\put(3.0900,11.09){\special{em:lineto}}
\put(1.9790,10.48){\special{em:lineto}}
\put(0.8936,9.896){\special{em:lineto}}
\put(-0.169,9.328){\special{em:lineto}}
\put(-1.211,8.792){\special{em:lineto}}
\put(-2.236,8.264){\special{em:lineto}}
\put(-3.245,7.755){\special{em:lineto}} }
\put(60,35){\unitlength=1.2mm\special{em:linewidth 0.3pt}
\put(-36.755,-7.755){\special{em:moveto}}
\put(-37.764,-8.264){\special{em:lineto}}
\put(-38.789,-8.792){\special{em:lineto}}
\put(-39.831,-9.328){\special{em:lineto}}
\put(-40.894,-9.896){\special{em:lineto}}
\put(-41.979,-10.48){\special{em:lineto}}
\put(-43.090,-11.09){\special{em:lineto}}
\put(-44.230,-11.73){\special{em:lineto}}
\put(-45.405,-12.41){\special{em:lineto}}
\put(-46.618,-13.12){\special{em:lineto}}
\put(-47.877,-13.88){\special{em:lineto}}
\put(-49.190,-14.69){\special{em:lineto}}
\put(-50.567,-15.57){\special{em:lineto}}
\put(-52.024,-16.53){\special{em:lineto}}
\put(-53.578,-17.58){\special{em:lineto}}
\put(-55.257,-18.76){\special{em:lineto}}
\put(-57.098,-20.10){\special{em:lineto}}
\put(-60.000,-22.20){\special{em:lineto}} }
\put(60,35){\unitlength=1.2mm\special{em:linewidth 0.3pt}
\put(-36.644,-7.700){\special{em:moveto}}
\put(-36.592,-8.668){\special{em:lineto}}
\put(-36.538,-9.164){\special{em:lineto}}
\put(-36.454,-9.680){\special{em:lineto}}
\put(-36.350,-10.14){\special{em:lineto}}
\put(-36.222,-10.62){\special{em:lineto}}
\put(-36.072,-11.08){\special{em:lineto}}
\put(-35.892,-11.54){\special{em:lineto}}
\put(-35.691,-11.99){\special{em:lineto}}
\put(-35.461,-12.43){\special{em:lineto}}
\put(-35.209,-12.88){\special{em:lineto}}
\put(-34.927,-13.30){\special{em:lineto}}
\put(-34.620,-13.72){\special{em:lineto}}
\put(-34.282,-14.12){\special{em:lineto}}
\put(-33.915,-14.52){\special{em:lineto}}
\put(-33.515,-14.90){\special{em:lineto}}
\put(-33.080,-15.25){\special{em:lineto}}
\put(-32.608,-15.60){\special{em:lineto}}
\put(-32.094,-15.91){\special{em:lineto}}
\put(-31.532,-16.20){\special{em:lineto}}
\put(-30.920,-16.45){\special{em:lineto}}
\put(-30.241,-16.67){\special{em:lineto}}
\put(-29.486,-16.84){\special{em:lineto}}
\put(-28.637,-16.93){\special{em:lineto}}
\put(-27.658,-16.94){\special{em:lineto}}
\put(-26.492,-16.84){\special{em:lineto}}
\put(-25.951,-16.74){\special{em:lineto}}
\put(-25.013,-16.52){\special{em:lineto}}
\put(-23.824,-16.13){\special{em:lineto}}
\put(-22.742,-15.68){\special{em:lineto}}
\put(-21.980,-15.31){\special{em:lineto}}
\put(-20.548,-14.55){\special{em:lineto}}
\put(-19.453,-13.85){\special{em:lineto}}
\put(-18.020,-12.88){\special{em:lineto}}
\put(-17.259,-12.31){\special{em:lineto}}
\put(-16.176,-11.47){\special{em:lineto}}
\put(-14.986,-10.49){\special{em:lineto}}
\put(-14.048,-9.660){\special{em:lineto}}
\put(-13.507,-9.168){\special{em:lineto}}
\put(-12.341,-8.060){\special{em:lineto}}
\put(-11.362,-7.060){\special{em:lineto}}
\put(-10.513,-6.168){\special{em:lineto}}
\put(-9.7588,-5.340){\special{em:lineto}}
\put(-9.0808,-4.548){\special{em:lineto}}
\put(-8.4684,-3.800){\special{em:lineto}}
\put(-7.9056,-3.100){\special{em:lineto}}
\put(-7.3924,-2.408){\special{em:lineto}}
\put(-6.9196,-1.748){\special{em:lineto}}
\put(-6.4840,-1.100){\special{em:lineto}}
\put(-6.0848,-0.488){\special{em:lineto}}
\put(-5.7180,0.1200){\special{em:lineto}}
\put(-5.3788,0.7200){\special{em:lineto}}
\put(-5.0736,1.3000){\special{em:lineto}}
\put(-4.7916,1.8720){\special{em:lineto}}
\put(-4.5384,2.4200){\special{em:lineto}}
\put(-4.3088,2.9920){\special{em:lineto}}
\put(-4.1092,3.5400){\special{em:lineto}}
\put(-3.9280,4.0720){\special{em:lineto}}
\put(-3.7788,4.6200){\special{em:lineto}}
\put(-3.6508,5.1400){\special{em:lineto}}
\put(-3.5472,5.6800){\special{em:lineto}}
\put(-3.4620,6.1600){\special{em:lineto}}
\put(-3.4080,6.6640){\special{em:lineto}}
\put(-3.3556,7.1040){\special{em:lineto}}
\put(-3.3556,7.7000){\special{em:lineto}} }
\put(60,35){\unitlength=1.2mm\special{em:linewidth 0.3pt}
\put(21.347,21.85){\special{em:moveto}}
\put(17.318,18.32){\special{em:lineto}}
\put(14.726,16.23){\special{em:lineto}}
\put(12.722,14.72){\special{em:lineto}}
\put(11.041,13.54){\special{em:lineto}}
\put(9.5620,12.56){\special{em:lineto}}
\put(8.2230,11.73){\special{em:lineto}}
\put(6.9860,10.99){\special{em:lineto}}
\put(5.8280,10.33){\special{em:lineto}}
\put(4.7295,9.730){\special{em:lineto}}
\put(3.6810,9.180){\special{em:lineto}}
\put(2.7730,8.675){\special{em:lineto}} }
\put(60,35){\unitlength=1.2mm\special{em:linewidth 0.3pt}
\put(-42.900,-8.675){\special{em:moveto}}
\put(-43.681,-9.180){\special{em:lineto}}
\put(-44.730,-9.730){\special{em:lineto}}
\put(-45.828,-10.33){\special{em:lineto}}
\put(-46.986,-10.99){\special{em:lineto}}
\put(-48.223,-11.73){\special{em:lineto}}
\put(-49.562,-12.56){\special{em:lineto}}
\put(-51.041,-13.54){\special{em:lineto}}
\put(-52.722,-14.72){\special{em:lineto}}
\put(-54.726,-16.23){\special{em:lineto}}
\put(-57.318,-18.32){\special{em:lineto}}
\put(-60.000,-20.55){\special{em:lineto}} }
\put(60,35){\unitlength=1.2mm\special{em:linewidth 0.3pt}
\put(-42.904,-8.8000){\special{em:moveto}}
\put(-42.904,-9.8224){\special{em:lineto}}
\put(-42.768,-10.859){\special{em:lineto}}
\put(-42.540,-11.918){\special{em:lineto}}
\put(-42.224,-12.993){\special{em:lineto}}
\put(-41.828,-14.084){\special{em:lineto}}
\put(-41.360,-15.190){\special{em:lineto}}
\put(-40.820,-16.299){\special{em:lineto}}
\put(-40.220,-17.412){\special{em:lineto}}
\put(-39.558,-18.520){\special{em:lineto}}
\put(-38.836,-19.612){\special{em:lineto}}
\put(-38.048,-20.678){\special{em:lineto}}
\put(-37.190,-21.704){\special{em:lineto}}
\put(-36.245,-22.670){\special{em:lineto}}
\put(-35.192,-23.549){\special{em:lineto}}
\put(-33.992,-24.300){\special{em:lineto}}
\put(-32.574,-24.850){\special{em:lineto}}
\put(-31.566,-25.040){\special{em:lineto}}
\put(-30.799,-25.064){\special{em:lineto}}
\put(-29.434,-24.890){\special{em:lineto}}
\put(-28.307,-24.597){\special{em:lineto}}
\put(-27.644,-24.350){\special{em:lineto}}
\put(-25.976,-23.53){\special{em:lineto}}
\put(-24.841,-22.870){\special{em:lineto}}
\put(-22.512,-21.230){\special{em:lineto}}
\put(-20.850,-19.860){\special{em:lineto}}
\put(-19.153,-18.42){\special{em:lineto}}
\put(-16.776,-16.270){\special{em:lineto}}
\put(-15.159,-14.750){\special{em:lineto}}
\put(-14.022,-13.660){\special{em:lineto}}
\put(-12.358,-12.070){\special{em:lineto}}
\put(-11.102,-10.820){\special{em:lineto}}
\put(-9.2012,-8.9368){\special{em:lineto}}
\put(-7.4260,-7.1500){\special{em:lineto}}
\put(-6.0064,-5.6988){\special{em:lineto}}
\put(-4.8068,-4.4500){\special{em:lineto}}
\put(-3.7528,-3.3280){\special{em:lineto}}
\put(-2.8080,-2.2940){\special{em:lineto}}
\put(-1.9504,-1.3208){\special{em:lineto}}
\put(-1.1644,-0.3880){\special{em:lineto}}
\put(-0.4400,0.52120){\special{em:lineto}}
\put(0.22160,1.41400){\special{em:lineto}}
\put(0.82120,2.30000){\special{em:lineto}}
\put(1.35960,3.18920){\special{em:lineto}}
\put(1.82880,4.08520){\special{em:lineto}}
\put(2.22400,4.99320){\special{em:lineto}}
\put(2.53920,5.91720){\special{em:lineto}}
\put(2.77000,6.86120){\special{em:lineto}}
\put(2.90520,8.82360){\special{em:lineto}} } \end{picture}}

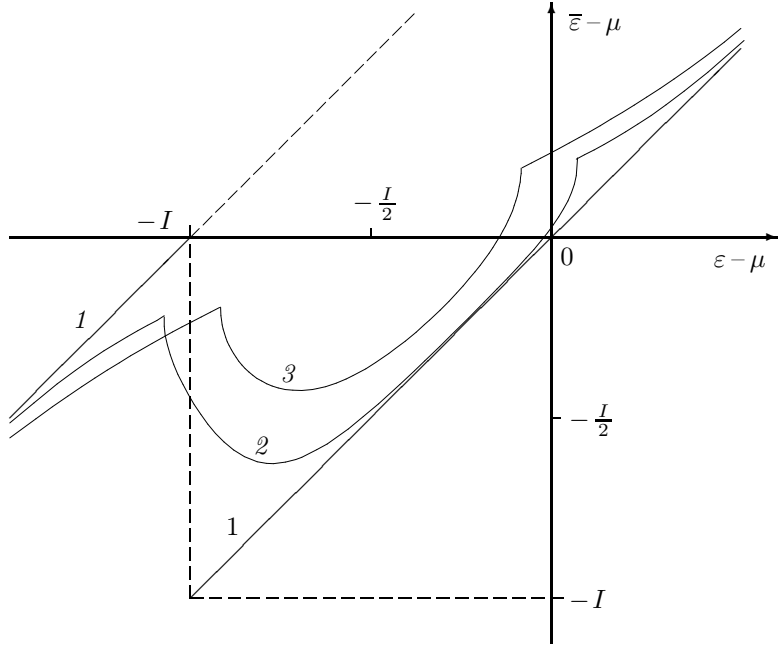
\vskip 15mm\centerline{\it Fig. 11. Mean electron
$\overline\varepsilon$ energy dependence of its $\varepsilon$
kinetic energy } \centerline {\it at various temperature values
$\tau$: 1 -- $\tau=0$, 2 -- $\tau=0,5$, 3 -- $\tau=0,8$. }

\vskip 5mm \centerline{\bf 12. Conclusion } \vskip 2mm

\par Thus, in scope of the theory discussed the microscopic
superconductivity is due to anisotropic wave vector electron
distribution. Normally, anisotropy is caused by electron repulsion
effect in the $\bf k$ and $-\hh\bf k$ wave vectors states. If
applied to the matter of more complex nature, model Hamiltonian is
formulated as follows: $$
\varepsilon_{\fbk\fbk^\pr}=I\hs\delta_{\fbk +\fbk^\pr}-
J\hs\delta_{\fbk-\fbk^\pr}\hs , $$ where $J$ is  attractive energy
of two electrons in the conditions equal to wave vectors ${\bf
k}={\bf k^\pr}$. This Hamiltonian is calculated in [9].

\newpage

\vskip 5mm \centerline{\bf Refereces } \vskip 3mm

\hskip-13pt[1] P.W.Anderson, Local Momentns and Localized States.
Nobel Lecture, 8 December 1977. \vskip 1mm

\hskip-13pt[2] P.W.Anderson, UFN, 1979, v.127, №.1, p.19. \vskip 1mm

\hskip-13pt[3] Yu.I.Sirotin, M.P.Shaskolskaya, Basic
Crystallophysics, M.: Nauka, 1979. \vskip 1mm

\hskip-13pt[4] B.V.Bondarev, N.P.Kalashnikov, G.G.Spirin, General
Physics Course, v.3, M.: Higher School, \linebreak\phantom{x}\hskip
3.5mm 2003. \vskip 1mm

\hskip-13pt[5] B.V.Bondarev, Vestinik MAI, 1996, v.3, №.2, p.56.
\vskip 1mm

\hskip-13pt[6] H.Kamerlingh Onnes, Comm. Phys. Leb. Univ. Leiden,
1911, №122, p.13. \vskip 1mm

\hskip-13pt[7] J.Bardeen, L.N.Cooper, J.R.Schrieffer, Phys.Rev.,
1957, v.106, №1, p.162; 1957, v.108, №5, \linebreak\phantom{x}\hskip
4mm p.1175. \vskip 1mm

\hskip-13pt[8] J.Schiffer, Superconductivity Theory, M.: Nauka,
1970. \vskip 1mm

\hskip-13pt[9] B.V.Bondarev, Density Matrix Method in Quantum Theory
of Cooperative Process, M.: Sputnik, \linebreak\phantom{x} \hskip
3mm 2001. \vskip 1mm

\end{document}